\documentclass{ejm_modified}
\usepackage{graphicx}
\usepackage{amsmath, amssymb}
\usepackage[inactive,tightpage]{preview}

\usepackage{microtype}
\usepackage[normalem]{ulem}
\usepackage{mathtools}

\usepackage{esvect}
\usepackage{fancyhdr}

\usepackage{enumitem}

\pdfmapfile{+rsfso.map}
\DeclareSymbolFont{rsfso}{U}{rsfso}{m}{n}
\DeclareSymbolFontAlphabet{\mathscr}{rsfso}

\usepackage{pdflscape}
\usepackage{afterpage}
\usepackage{flafter}
\usepackage{rotating}
\usepackage{fancyhdr}
\fancypagestyle{lscape}{%
\fancyhf{} 
\fancyfoot[LE]{}
\fancyfoot[LO] {}
 
}

\usepackage{bm}
\usepackage{microtype}

\usepackage[pdfauthor={Philippe H. Trinh}]{hyperref}
\usepackage[usenames, dvipsnames]{xcolor}
\hypersetup{colorlinks=true,linkcolor=MidnightBlue,citecolor=MidnightBlue}

\usepackage{natbib}
\usepackage{array}
\usepackage{booktabs}
\usepackage{multirow}

\usepackage[small]{caption}
\usepackage{subcaption}
\usepackage{tabularx}
\newcolumntype{Y}{>{\centering\arraybackslash}X}

\graphicspath{ {images/} }

\usepackage[percent]{overpic}
\usepackage[ruled,linesnumbered]{algorithm2e}
\usepackage{float}

\usepackage{tikz}
\usepackage{calrsfs}

\newcommand*{\ep}{\epsilon}
\newcommand*{\F}{F}
\newcommand*{\E}{\mathcal{E}}
\newcommand*{\Ehigh}{E_{\text{hw}}}
\renewcommand*{\i}{\mathrm{i}}

\newcommand*{\de}{\operatorname{d\!}{}} 

\newcommand{\G}[2]{G_{{#1}\to{#2}}}

\usepackage{mathabx}
\usepackage{esint} 
\def\Xint#1{\mathchoice
   {\XXint\displaystyle\textstyle{#1}}%
   {\XXint\textstyle\scriptstyle{#1}}%
   {\XXint\scriptstyle\scriptscriptstyle{#1}}%
   {\XXint\scriptscriptstyle\scriptscriptstyle{#1}}%
   \!\int}
\def\XXint#1#2#3{{\setbox0=\hbox{$#1{#2#3}{\int}$}
     \vcenter{\hbox{$#2#3$}}\kern-.5\wd0}}

\def\YYint#1#2#3{{\setbox0=\hbox{$#1{#2#3}{\int}$}
     \vcenter{\hbox{\scalebox{1}[-1]{$#2#3$}}}\kern-.5\wd0}}

\def\dashint{\Xint-}

\shorttitle{Parasitic gravity-capillary waves}
\shortauthor{Shelton, Milewski, and Trinh}

\title{On the structure of steady parasitic gravity-capillary waves in the \\ small surface tension limit}

\author{Josh Shelton \footnotemark, Paul Milewski,  \and Philippe H. Trinh \footnotemark}

\affiliation{Department of Mathematical Sciences, University of Bath, Bath BA2 7AY, UK}

\pubyear{XXXX}
\volume{YYY}
\pagerange{xxx--xxx}
\date{\today}

\begin{document}
\maketitle
\begin{abstract}
When surface tension is included in the classical formulation of a steadily-travelling gravity wave (a Stokes wave), it is possible to obtain solutions that exhibit parasitic ripples: small capillary waves riding on the surface of steep gravity waves. However, it is not clear whether the singular small-surface-tension limit is well-posed. That is, is it possible for an appropriate travelling gravity-capillary wave to be continuously deformed to the classic Stokes wave in the limit of vanishing surface tension? The work of Chen \& Saffman [Stud. Appl. Math. 1980, \textbf{62} (1) 1--21] had suggested smooth continuation was not possible, while the work of Schwartz \& Vanden-Broeck [J. Fluid Mech. 1979, \textbf{95} (1) 119--139] presented an incomplete bifurcation diagram of the nonlinear numerical solutions. In this paper, we numerically explore the low surface tension limit of the steep gravity-capillary travelling-wave problem. Our results allow for a classification of the bifurcation structure that arises, and serve to unify a number of previous numerical studies. Crucially, we demonstrate that different choices of solution amplitude can lead to subtle restrictions on the continuation procedure. When wave energy is used as a continuation parameter, solution branches can be continuously deformed to the zero surface tension limit of a travelling Stokes wave. 
\end{abstract}

\newcommand\blfootnote[1]{%
  \begingroup
  \renewcommand\thefootnote{}\footnote{#1}%
  \addtocounter{footnote}{-1}%
  \endgroup
}

\blfootnote{$\dagger$ Email address for correspondence: j.shelton@bath.ac.uk}
\blfootnote{$\ddagger~$Email address for correspondence: p.trinh@bath.ac.uk}

\author{Shelton, Milewski, \and Trinh}
\title{Parasitic gravity-capillary waves}

\section{Introduction}
\label{sec:Intro}


%
\noindent In this paper, we consider the two-dimensional formulation of a travelling gravity-capillary wave on a fluid of infinite depth. When posed in a travelling frame, the steady non-dimensionalised problem is to determine a velocity potential, $\phi(x, y)$, which is harmonic in a periodic domain, $-\tfrac{1}{2} \leq x \leq \tfrac{1}{2}$ and $-\infty < y \leq \zeta(x)$. 
On the unknown free surface, $y = \zeta(x)$, Bernoulli's condition requires that 
\begin{equation}\label{eq:PotFlow}
  \frac{F^2}{2} |\nabla \phi|^2 + \zeta - B \kappa = \text{const.}
\end{equation}
Here, the Froude number, $F$, characterises the balance between inertia and gravity, and is proportional to the wave-speed.
The inverse-Bond number, $B$, characterises the balance between gravity and surface tension; with this latter effect depending on the surface curvature, $\kappa$. Typically, a wave energy or amplitude parameter, $\E$, is fixed and prescribes the degree of nonlinearity. Solutions are then characterised by bifurcation curves in $(B,F)$ or $(B,F,\E)$-solution space. The small surface tension limit corresponds to $B \to 0$.

Extensive results are known for the case with $B=0$ when surface tension is neglected, and this originates from the seminal work of \cite{stokes_1847a_on_the}; cf. the reviews by \cite{Okamoto:2001aa,Toland:1996aa}. 
Intuitively, we might expect that the inclusion of a small amount of surface tension results in a small change in the profile of the pure gravity wave. 
However, since the limit of $B \to 0$ is singularly perturbed, this is not necessarily the case, and it is known that the introduction of surface tension has a significant impact on the existence and uniqueness of solutions, their bifurcations, and their profiles.

The goal of this paper is to present a numerical study of nonlinear solutions in the singular limit of $B \to 0$, for which we know one solution to be the Stokes wave. 
We demonstrate the numerical existence of a cohesive structure of branches of solutions existing under this limit.
Importantly this suggests that, with fixed wave energy, $\E$, only one of a family of solutions approaches the classical Stokes wave as $B \to 0$.

We firstly discuss the analytical and numerical difficulties of the $B \to 0$ limit.

\subsection{Longuet-Higgins and parasitic ripples}\label{sec:L-H}

\begin{figure}
\includegraphics[scale=1.2]{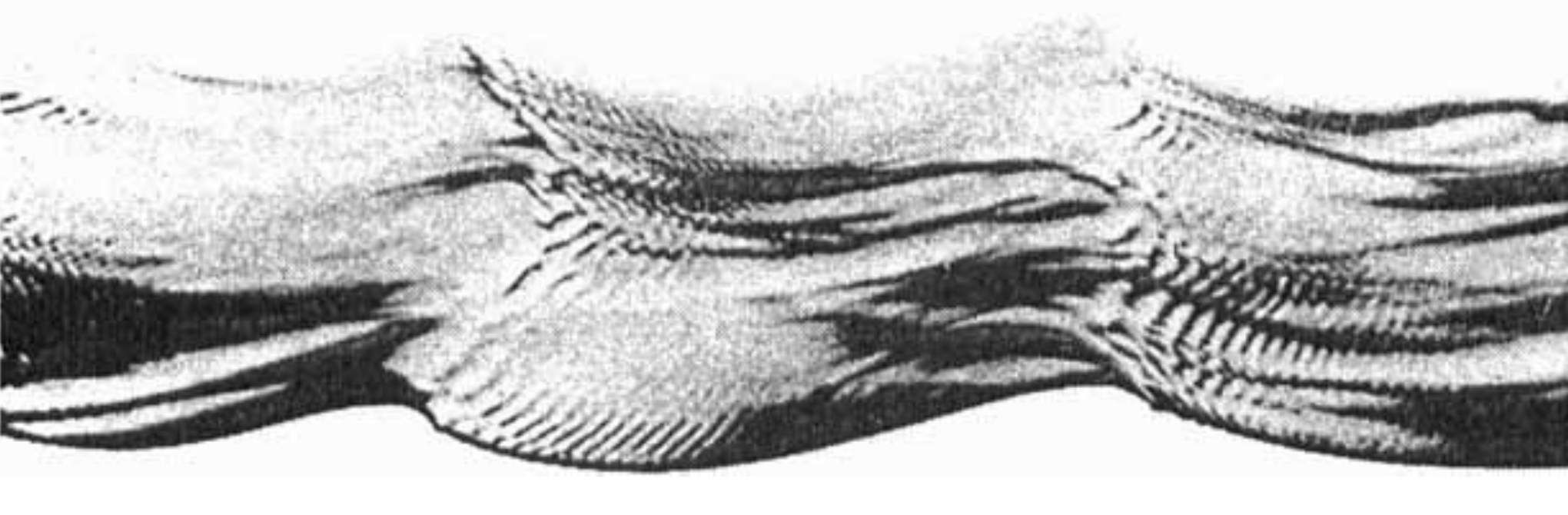}
\caption{\label{fig:ripples} Experimental picture showing parasitic ripples located near the crests of steep gravity waves [from \cite{ebuchi1987fine}].}
\end{figure}

\noindent It is well-known observationally that under the action of both gravity and surface tension, ripples of small wavelength form on the forward face of a propagating wave. As shown by the experimental results of \cite{cox1958measurements} and \cite{ebuchi1987fine} for instance, the amplitude of these {\it{parasitic capillary ripples}} increases when the overall amplitude of the wave (measured by crest to trough displacement) increases. An example of such parasitic ripples, as photographed by \cite{ebuchi1987fine}, is shown in figure~\ref{fig:ripples}, where it is seen that these ripples are asymmetric about the wave crest and unsteady in the frame of the propagating wave.

In the simplest steady framework, one assumes that the parasitic ripples are fixed to the same travelling frame of reference as the underlying gravity wave. A steady asymmetric theory was proposed by \cite{longuet-higgins_1963_the_generation} to describe the form of these waves. However, his method was asymptotically inconsistent, leading to poor agreement with the experimental results of \cite{perlin1993parasitic}. We shall provide a preliminary discussion of these issues in \S\ref{sec:Discussion}. 

Nevertheless, it remains an open question as to whether parasitic capillary ripples similar to those shown in figure~\ref{fig:ripples} may be found as either symmetric or asymmetric solutions of the steady framework of equation \eqref{eq:PotFlow}.
In this work, we present clear numerical evidence that steady-symmetric parasitic ripples do exist within the solution space of the classical potential framework in the $B \to 0$ limit.

\subsection{Schwartz \& Vanden-Broeck and the complexity of $(B,F)$-space}

\begin{figure}
\begin{centering}
\includegraphics[scale=1]{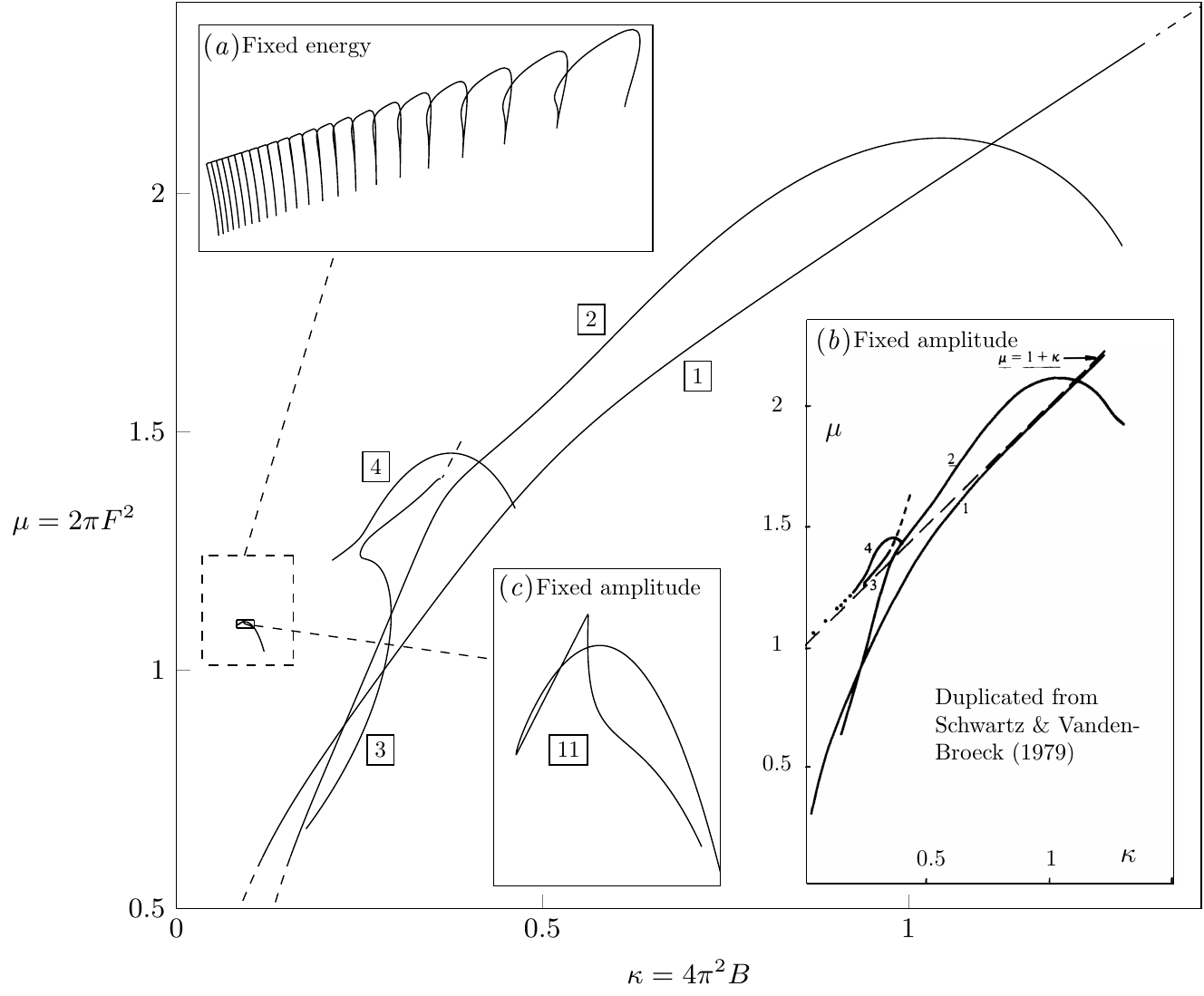}
\end{centering}
\caption{\label{fig:3d} Our solutions of fixed energy from \S\ref{sec:Results} are shown in ({\it{a}}). A duplication of the fixed amplitude results of Schwartz and Vanden-Broeck is displayed in ({\it{b}}). These branches of fixed amplitude are shown in the $(\kappa,\mu)$ plane, where the boxed type number inditcates the number of observed `dimples' or inflexion points on a (half) wave profile. In ({\it{c}}) we show the type-11 fixed amplitude branch.}
\end{figure}

\noindent In their seminal work \cite{schwartz1979numerical} developed a numerical scheme using a series truncation method to compute periodic gravity-capillary waves of the exact nonlinear equations. Imposing symmetry at $x=0$ and an amplitude condition on the crest-to-trough displacement, they presented a preliminary classification of solutions in $(B,F)$-space of types 1, 2, 3, and 4. Each type number was associated with a distinct branch of solutions, and corresponded to the number of observed `dimples' or inflexion points on a (half) wave profile. 

A reproduction of their original bifurcation diagram, which is computed at fixed crest-to-trough amplitude, is shown in subfigure~\ref{fig:3d}$(b)$. 
Our intention in reproducing this figure is to convince the reader that indeed the bifurcation space of the gravity-capillary problem is certainly non-trivial, and it is difficult to observe any clear structure. 
We also show the computed \cite{schwartz1979numerical} bifurcation curves in figure~\ref{fig:3d} alongside our solutions of fixed energy.



\begin{figure}
\begin{centering}
\includegraphics[scale=1]{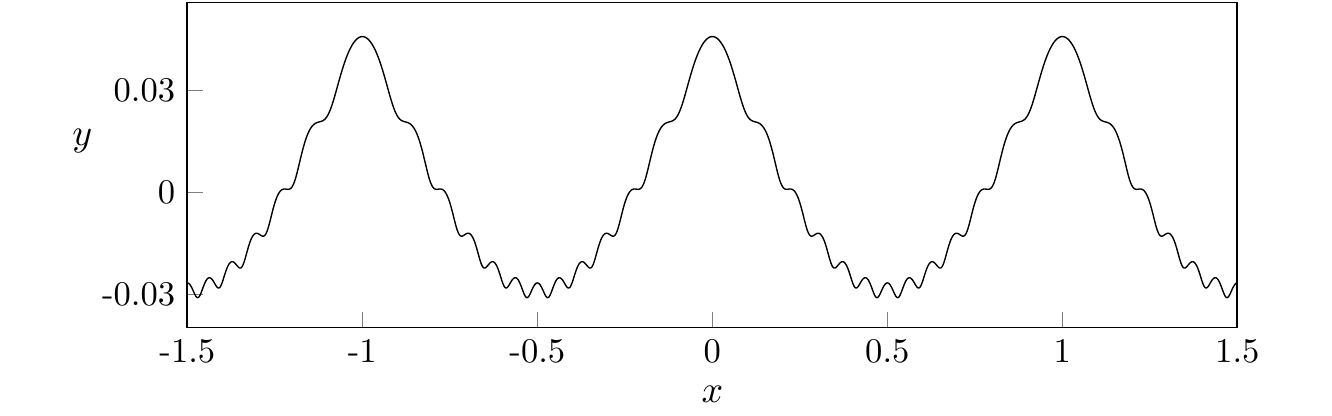}
\end{centering}
\caption{\label{fig:ripples2} A numerical solution of equations \eqref{eq:MainEq} and \eqref{eq:MainHarm} is displayed in physical $(x,y)$ space, with nondimensional parameters $F=0.4299$, $B=0.002270$, and energy $\E=0.3804$. This has been computed using the scheme described in \S\ref{sec:num}. The periodic solution has been repeated three times.}
\end{figure}

One of their solutions [\cite{schwartz1979numerical} fig.~10] is of particular interest in the context of parasitic ripples. This profile, similar to that shown in figure~\ref{fig:ripples2}, appears to contain small-scale capillary ripples as a perturbation to the main Stokes wave. This is one of the types of solution that we will be expanding upon in this work.
Note in addition that the type 1 to 4 branches, as shown in their figure, have a non-trivial and unstructured shape in the bifurcation diagram; it is not obvious if a more consistent pattern emerges upon increasing the type number, or whether these solution curves can be taken as $B \to 0$. We shall explain the reason for these issues in this work.

Later, in seeking to compare new experimental data with the previous analytical approximations of \cite{longuet-higgins_1963_the_generation} and numerical solutions of \cite{schwartz1979numerical}, \cite{perlin1993parasitic} made extensive remarks on the challenges of navigating the solution space of the full nonlinear problem, noting that \emph{``there is no known method for determining the number of solutions to the numerical formulation\ldots''} (p.~618). Indeed, they state that (p.~598)
\begin{quotation}\noindent\emph{Surprisingly little information is available on these waves of disparate scales, presumably due to the analytical/numerical, as well as experimental, difficulties involved.}
\end{quotation}


\subsection{Chen \& Saffman and the impossibility of the $B \to 0$ limit}

\noindent Nearly in parallel with the work by \cite{schwartz1979numerical}, \cite{chen1979steady,chen1980numerical,chen1980steady} produced a series of works where they examined the Stokes wave problem, largely from the perspective of weakly nonlinear theory (and its numerical consequences on the full nonlinear problem). 

In \cite{chen1979steady} they considered weakly nonlinear solutions of equation \eqref{eq:PotFlow} in powers of a small wave-amplitude, $\epsilon$. Expressing the solution, $y=\zeta(x)$, as a Fourier series, this permits analytical solutions for the Fourier coefficients, $A_n$. They discovered that in fixing the point of symmetry of the wave-profile to be at $x=0$, the branches of solutions in the $(\kappa,A_n)$ bifurcation space (where $\kappa=4\pi^2 B$) are discontinuous either side of the point $\kappa=1/n$. Due to this discontinuity and the analytical criterion, they commented (p.~204):
\begin{quotation}\noindent\emph{The gravity wave ($\kappa=0$) is therefore a singular limit which cannot be reached smoothly by applying the limit $\kappa \to 0$ to a gravity capillary wave.}
\end{quotation}

\noindent In our work, we will note how this statement is misleading since their non-smoothness is a consequence of their initial assumption of a fixed point of symmetry at $x=0$.
We demonstrate this in \S\ref{sec:PastworkSym}, noting that this is due to the presence of a {\it symmetry shifting bifurcation}.
Thus, if their enforced symmetry at $x=0$ were to be relaxed, the branches of solutions in $(\kappa, A_n)$-space either side of the point $\kappa=1/n$ would be continuous.

 In a second work by \cite{chen1980steady}, the numerical solution space was explored for waves of finite amplitude. 
Their choice of amplitude was a linear combination of Fourier coefficients, typically chosen to be that of the fundamental mode, $A_1$, or the $n$th mode $A_n$.
Nonlinear solutions were computed. However, the authors were unable to connect branches of solutions in the resulting bifurcation diagram, from which they concluded:
\begin{quotation} \noindent \emph{These results confirm the impossibility of going continuously from a pure capillary-gravity wave to a gravity wave by letting $\kappa \to 0$.}   [\cite{chen1980steady}]
\end{quotation}

\noindent We later note in \S\ref{sec:PastworkAmp} that if the wave-energy instead is fixed as an amplitude parameter then the continuous set of solutions as $B \to 0$, discovered within this paper, bifurcate from solutions with fundamental mode $A_1=0$. 
Hence if $A_1$ is fixed to be a non-zero constant, as in the numerical work of \cite{chen1980steady}, this bifurcation point would remain undiscovered. 
We shall conclude that in order to achieve a smooth transition as $B \to 0$, the first Fourier coefficient should not be fixed.

\subsection{Outline of the paper}

\noindent  In this work we shall consider the numerical behaviour of steady symmetric parasitic ripples for small values of the Bond number, $B$.
Starting in \S\ref{sec:MF}, we introduce the governing equations for the gravity-capillary wave problem, which we transform to depend on the velocity potential, $\phi$, alone.
In \S\ref{sec:Linear}, the well known linear solutions are derived. 
These form a starting point for our numerical method of \S\ref{sec:num}.
Solutions are presented in \S\ref{sec:Results}, which we use to demonstrate that as $B \to 0$ for fixed energy, $\E$, this bifurcation structure appears to form a countably infinite number of smooth branches of solutions in the $(B,F)$ plane. 
Each smooth branch forms a ``finger'' in the solution space, which is connected continuously to the proceeding branch.
These branches then accumulate in the limit of $B \to 0$ such that solutions are conjectured to exist in an $O(1)$ interval in the Froude number, $F$.
These smooth branches of solutions are connected at the point where they bifurcate from a wave with smaller fundamental wavelength, resulting in numerical evidence for the $B \to 0$ limit of the steep gravity-capillary wave problem having a continuous set of solutions. 

This allows us to show why previous authors have failed to reveal this underlying structure, which we comment upon in \S\ref{sec:Pastwork} and \S\ref{sec:Conclusions}.
Lastly, in \S\ref{sec:Discussion}, we comment upon the asymptotic properties as $B \to 0$ of our discovered solutions, and how this uncovered structure is likely to be present in other numerical problems for small values of the Bond number, $B$.


\section{Mathematical formulation} \label{sec:MF}

\noindent Consider two-dimensional free surface flow of an inviscid, irrotational, and incompressible fluid of infinite depth. The velocity potential, $\phi$, is defined by $\mathbf{u} = \nabla \phi$. We assume that, in the lab frame, the free surface, $y = \zeta(x, t)$, is periodic in $x$ with wavelength $L_\lambda$, and moves to the right with wavespeed $c$. We non-dimensionalise with unit length, $L_\lambda$, and velocity, $c$. We consider steady travelling-wave solutions by introducing a subflow of unit horizontal velocity in the opposite direction of wave propagation. This negates the movement of the free-surface. Then $\zeta_t = 0 = \phi_t$ yields the steady governing equations [compare to \emph{e.g.} \citealt[eqns (2.48)--(2.55)]{vanden2010gravity}]
\begin{subequations} \label{eq:MF1}
\begin{align}
\phi_{xx}+\phi_{yy}=0& \ \ \ \ \text{for} \ \ y \leq \zeta, \label{eq:laplace} \\
\phi_{y}=\zeta_{x}\phi_{x}& \ \ \ \ \text{at} \ \   \ y=\zeta, \label{eq:kin}\\
\frac{F^2}{2}(\phi^{2}_{x}+\phi^{2}_{y})+y-B\frac{\zeta_{xx}}{(1+\zeta_x^2)^{\frac{3}{2}}}=\frac{F^2}{2}& \ \ \ \ \text{at} \ \   \ y=\zeta, \label{eq:dyn}\\
\phi_{y} \to 0 \quad \text{and} \quad \phi_{x} \to -1 & \ \ \ \ \text{as} \ \  \ y \to -\infty, \label{eq:deep}
\end{align}
\end{subequations}
for the travelling wave now in $x \in [-\tfrac{1}{2}, \tfrac{1}{2})$. Thus the system is governed by Laplace's equation \eqref{eq:laplace} within the fluid, kinematic and dynamic conditions on the free surface \eqref{eq:kin} and \eqref{eq:dyn}, and uniform flow conditions in the deep-water limit \eqref{eq:deep}. The horizontal velocity condition \eqref{eq:deep}, our subflow, indicates a uniform flow moving towards the left. The spatial subscripts in \eqref{eq:MF1} correspond to partial differentiation.

The non-dimensional Froude ($F$) and inverse-Bond ($B$) numbers are given by 
\begin{equation}
\label{eq:FB}
F=c/\sqrt{gL_\lambda} \qquad \text{and} \qquad B=\sigma/\rho g L_\lambda^2,
\end{equation}
where $g$ is the gravitational constant, $\rho$ is the fluid density, and $\sigma$ is the coefficient of surface tension.  

\emph{Remark on terminology:} note that in the mathematical formulation above, we have non-dimensionalised lengths by a fixed physical wavelength, $L_{\lambda}$, and hence we shall seek solutions that are $1$-periodic in the non-dimensional travelling frame. However, these solutions may have a smaller wavelength which is less than unity. We thus define $\lambda$ to be the non-dimensional \emph{fundamental wavelength} (the smallest such wavelength). Moreover in this work, we shall refer to a wave with fundamental wavenumber $k = 1/\lambda$ as a \emph{pure} $k$-wave. Thus a pure $k$-wave has a dimensional wavelength of $\lambda L_\lambda$. 



\subsection{The conformal mapping to the $(\phi,\psi)$ plane.} \label{sec:conformal}

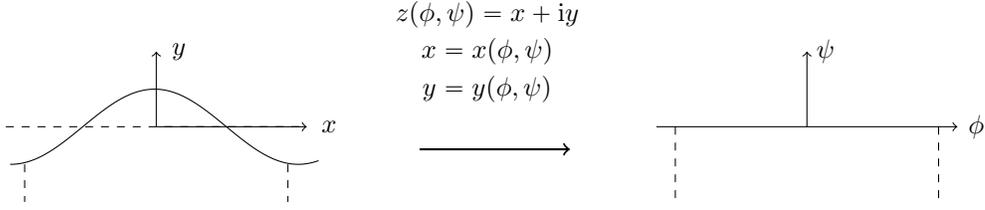
\begin{figure}
\centering
\begin{tikzpicture}
\draw [yshift=0][xshift=50][variable=\t, domain=1.9:6, samples=100, smooth] plot (\t, {0.5*cos(30*pi*\t )});
\draw (12.25,0)[->] -- (16.25,0);
\draw (14.25,0)[->] -- (14.25,1);
\draw (12.5,0)[dashed] -- (12.5,-1);
\draw (16,0)[dashed] -- (16,-1);
\draw (16.5,0) node {$\phi$};
\draw (14.5,1) node {$\psi$};
\draw (3.6,0)[dashed] -- (7.6,0);
\draw (5.6,0)[->] -- (7.6,0);
\draw (5.6,0)[->] -- (5.6,1);
\draw (7.9,0) node {$x$};
\draw (5.9,1) node {$y$};
\draw (3.85,-0.5)[dashed] -- (3.85,-1);
\draw (7.35,-0.5)[dashed] -- (7.35,-1);
\draw (10,1.5) node {$z(\phi,\psi)=x+{\mathrm{i}}y$};
\draw (10,1) node {$x=x(\phi,\psi)$};
\draw (10,0.5) node {$y=y(\phi,\psi)$};
\draw (9.1,-0.3)[->, thick] -- (11.1,-0.3);
\end{tikzpicture}
\caption{\label{fig:conf}  The conformal mapping from $(x,y)$ to $(\phi,\psi)$ is shown.}
\end{figure}

We now formulate the governing equations \eqref{eq:MF1} in the potential $(\phi,\psi)$-plane, as shown in figure~\ref{fig:conf}. We assume that the free-surface is located along $\psi = 0$, and introduce the notation of $X$ and $Y$ for the fluid quantities evaluated on the free surface. Thus
\begin{equation} \label{eq:Phi1}
X(\phi)\equiv x(\phi,0) \quad \text{and} \quad 
Y(\phi)\equiv \zeta(x(\phi,0)). 
\end{equation}
We may now obtain expressions for the surface derivative and curvature by differentiating \eqref{eq:Phi1}. This yields
\begin{equation}\label{eq:Phi5}
\zeta_{x}=\frac{Y_{\phi}}{X_{\phi}} \quad \text{and} \quad
\zeta_{x x}=\frac{X_{\phi}Y_{\phi \phi}-Y_{\phi}X_{\phi \phi}}{X_{\phi}^3}. 
\end{equation}
We now seek to re-write the kinematic~\eqref{eq:kin} and dynamic~\eqref{eq:dyn} boundary conditions on the surface in terms of the conformal variables $X$ and $Y$. First, the velocities, $\phi_x$ and $\phi_y$, may be inverted, yielding
\begin{equation} \label{eq:Phi2}
  \phi_x = \frac{x_{\phi}}{x_{\phi}^2+y_{\phi}^2} \quad \text{and} \quad  \phi_y = \frac{y_{\phi}}{x_{\phi}^2+y_{\phi}^2}.
\end{equation}

Finally, substitution of \eqref{eq:Phi2} for $\phi_x$ and $\phi_y$, \eqref{eq:Phi5} for $\zeta_x$ and $\zeta_{xx}$, and $Y(\phi)=\zeta(x(\phi,0))$ into Bernoulli's equation \eqref{eq:dyn} and setting $\psi=0$ yields our governing equation [compare with eqn. (6.12) of \cite{vanden2010gravity}]
\begin{equation}
\label{eq:PhiBern}
\frac{F^2}{2J}+Y+B\frac{(Y_{\phi}X_{\phi \phi}-X_{\phi}Y_{\phi \phi})}{J^{\frac{3}{2}}}=\frac{F^2}{2}.
\end{equation}
Above we have introduced the surface Jacobian, $J$, via
\begin{equation}
  J(\phi) = X_{\phi}^2+Y_{\phi}^2.
\end{equation}
Note that in the conformal formulation, the kinematic condition \eqref{eq:kin} can be verified to be satisfied identically once \eqref{eq:Phi5} and \eqref{eq:Phi2} are used on the streamline $\psi = 0$. 

In addition to Bernoulli's equation \eqref{eq:PhiBern}, in order to close the system, we require a harmonic relationship between $X$ and $Y$. Note that within the fluid, $y(\phi, \psi)$ can be written as a Fourier series of the form
\begin{equation}\label{eq:yFourier}  
y(\phi,\psi)=\psi+A_0+\sum_{n=1}^{\infty}e^{2n\pi \psi}\bigg[A_n\cos{(2n\pi \phi)}+B_n\sin{(2n\pi \phi)} \bigg],
\end{equation}
where $A_n$ and $B_n$ are real-valued for all $n$. Indeed, the above ansatz satisfies $y_{\phi\phi} + y_{\psi\psi} = 0$ along with the depth condition $y \sim \psi$ as $\psi \to -\infty$. 

We define the Hilbert transform on $Y$ by 
\begin{equation} \label{eq:Hilb_pre}
\mathscr{H}[Y](\phi^{\prime})=\dashint_{-\infty}^{\infty} \frac{Y(\phi)}{\phi-\phi^{'}} \, \de{\phi},
\end{equation}
where the integral is of principal-value type. Then by the assumed periodicity of the solution, this implies that 
\begin{equation} \label{eq:Hilb}
\mathscr{H}[Y](\phi^{\prime}) = \dashint_{-1/2}^{1/2} Y(\phi)\cot{[\pi(\phi-\phi^{\prime})]} \de{\phi}.
\end{equation}

We can then verify that the individual Fourier modes can be related using $\mathscr{H}[\sin(2n\pi \phi)]=\cos(2n \pi \phi)$ and $\mathscr{H}[\cos(2n \pi \phi)]=-\sin(2n \pi \phi)$. From using the Cauchy-Riemann relations of $x_{\phi}=y_{\psi}$ and $x_{\psi}=-y_{\phi}$, we obtain the harmonic relationships between $X$ and $Y$ on the free surface via
\begin{equation} \label{eq:Harm}
X_{\phi}(\phi)=1-\mathscr{H}[Y_{\phi}(\phi)] \quad \text{and} \quad 
Y_{\phi}(\phi) = \mathscr{H}[X_{\phi}(\phi) - 1].
\end{equation}
A choice of any one of the relations in \eqref{eq:Harm}, combined with Bernoulli's equation \eqref{eq:PhiBern} allows $X$ and $Y$ to be solved. 

\subsection{The energy constraint} \label{sec:energy}

\noindent  In order to fully close the formulation, we shall impose an energy constraint on the solution, which can be viewed as equivalent to a measurement of the wave amplitude. 
We define the wave energy, $E$, to be
\begin{equation} \label{eq:Energy}
E=\frac{F^2}{2}\int_{-\frac{1}{2}}^{\frac{1}{2}}Y(X_{\phi}-1) \de \phi +B \int_{-\frac{1}{2}}^{\frac{1}{2}}(\sqrt{J}-X_{\phi})\de \phi +\frac{1}{2}\int_{-\frac{1}{2}}^{\frac{1}{2}}Y^2 X_{\phi} \de \phi,
\end{equation}
where the first integral on the right hand-side corresponds to the kinetic energy, the second to the capillary potential energy, and the third to the gravitational potential energy. The derivation of \eqref{eq:Energy} from the bulk energy is given in Appendix~\ref{sec:Energy}. 

For comparison purposes, it will convenient for us to re-scale the energy in \eqref{eq:Energy} by the energy of the highest (fundamental) gravity wave, $E_\text{hw}$. 
Thus we write 
\begin{equation}\label{eq:EnNorm}
  \E = \frac{E}{E_{\text{hw}}},
\end{equation}
where $E_\text{hw} \approx 0.00184$ (to 5 decimal places) is calculated using the numerical scheme of \S\ref{sec:num} applied to the pure gravity wave using $n=4096$ Fourier coefficients.

The choice of how to define an amplitude or energy condition for the wave is a subtle one. In this paper, we shall comment on the following three choices of amplitude:
\begin{equation} \label{eq:Amplitudes}
\mathscr{A}=\left\{\begin{aligned}
 &\text{$\E$} \\
  &\text{$A_1$} \\
 &\text{$Y(0)-Y(1/2)$}
\end{aligned}\right.
\qquad
\begin{aligned}
 &\text{[energy definition from \eqref{eq:Energy}]}, \\
 &\text{[first Fourier coefficient from \eqref{eq:yFourier}]}, \\
 &\text{[crest-to-trough displacement]}.
\end{aligned}
\end{equation}
The second choice of $A_\text{1}$, as used in \cite{chen1980steady}, designates the amplitude to be the first Fourier coefficient, while the third choice of $Y(0)-Y(1/2)$, as used by \cite{schwartz1979numerical}, is a sensible choice to measure the physical wave height of the fundamental Stokes wave.


Note that both definitions of amplitude, $A_{\text{1}}$ and $Y(0)-Y(1/2)$, have the problem that strongly nonlinear waves (as measured by a lack of decay in the Fourier coefficients) can occur, even at small amplitude values. This is particularly affected by the fact that gravity-capillary waves may take a variety of shapes beyond the simple fundamental wave considered by \cite{stokes_1847a_on_the}. Similar difficulties were encountered by \cite{chen1979steady,chen1980steady} who chose $\mathscr{A} = A_1$ but commented that:
\begin{quotation}\noindent \emph{We found from experience that none of these parameters were universally useful for describing the bifurcation phenomenon to be described in this work, and in fact we have been unable to construct a parameter which characterized the magnitude of the wave for all the phenomena in a satisfactory way.} [\cite{chen1979steady}]
\end{quotation}
It may be that using the energetic definition of amplitude with $\mathscr{A}=\E$ is the modification required to fix these issues; indeed within the context of our numerical investigations this does seem to be the case in the small surface-tension limit.

\section{Linear theory, Wilton ripples, and type $(n,m)$-waves} \label{sec:Linear}

\noindent  It will be useful for us to review linear solutions in the notation of \S\ref{sec:conformal}. The results of linear theory are found from the first two terms of a Stokes expansion in powers of a small amplitude parameter, $\epsilon$ [see \emph{e.g.} \cite[Sec.~2.4.2]{vanden2010gravity}]. Thus we shall consider equations \eqref{eq:MainEq} and \eqref{eq:MainHarm} and take $X\sim X_0+ \epsilon X_1$ and $Y \sim Y_0 + \epsilon Y_1$. Solving the resultant equations yields $X_0=\phi$ and $Y_0=0$ at leading order. At $O(\epsilon)$, we write $X_1$ and $Y_1$ as Fourier series and assume that the two solutions are respectively odd and even about $\phi=0$. This yields the necessary equation that
\begin{equation}
\label{eq:AB1}
\sum_{k=1}^{\infty}\bigl[ F^2(2k\pi)-1-(2k\pi)^2B \bigr] a_k \cos{(2k\pi \phi)}=0.
\end{equation}
In order to obtain non-trivial solutions, we require the linear dispersion relation of 
\begin{equation}
\label{eq:disp}
2k\pi F^2-1-4k^2\pi^2B=0,
\end{equation}
and obtain $X_1 = a_k \sin(2k \pi \phi)$ and $Y_1=a_k \cos{(2k \pi \phi)}$. Thus the linear solution, a pure--$k$ wave, is approximated by
\begin{equation} \label{eq:linearXY}
X \sim \phi + \ep\Bigl[ a_k \sin(2k \pi \phi) \Bigr] \quad \text{and} \quad
Y \sim  \ep \Bigl[ a_k \cos(2k \pi \phi) \Bigr],
\end{equation}
to the first two orders. Substitution into the energy expression \eqref{eq:EnNorm} yields
\begin{equation} \label{eq:linearEnergy}
  \E \sim \epsilon^2 \frac{ 2 K^2 \pi^2 B a_k^2}{E_{\text{hw}}}.
\end{equation}

The linear solution \eqref{eq:linearXY} was assumed to satisfy the single dispersion relation \eqref{eq:disp} for the $k$th Fourier mode only. Note that other solutions may be constructed that satisfy the dispersion relation for more than one mode. For instance, if the modes with $k=1$ and $k=n$ are assumed to be non-degenerate, then we require that both $2\pi F^2-1-4\pi^2B=0$ and $2n\pi F^2-1-4n^2\pi^2B=0$. This yields the so-called Wilton ripples predicted by \cite{wilton1915lxxii}, located wherever
\begin{equation} \label{eq:linearwiltonBF}
B_\text{wilton} = \frac{1}{4\pi^2 n} \quad \text{and} \quad
F_\text{wilton}^2 = \frac{(1+n)}{2\pi n},
\end{equation} 
with $n \in\mathbb{Z}^+$. The Wilton ripples are then given by 
\begin{equation} \label{eq:linearwilton}
X_1 = a_1 \sin(2 \pi \phi)+a_n \sin(2n \pi \phi)  \quad \text{and} \quad
Y_1 =  a_1 \cos(2 \pi \phi) +a_n \cos(2n \pi \phi) .
\end{equation} 
In the numerics of \S\ref{sec:num}, we shall often initialise the numerical continuation method with the linear solution \eqref{eq:linearXY} using a small value of $\ep a_k$. Crucially, since this linear solution is invalid near points~\eqref{eq:linearwiltonBF} we must ensure that our initial choice lies away from the critical numbers of $B_\text{wilton}$ and/or $F_\text{wilton}$. 

As introduced by \cite{chen1979steady}, linear solutions that consist of a combination of pure $n$- and $m$-waves, and with fundamental wavelengths of $\lambda=1/n$ and $1/m$ respectively, are described as a \emph{type (n,m)-wave}. Thus under this terminology, Wilton's solution in \eqref{eq:linearwiltonBF} is an example of a type $(1,n)$-wave. Our numerical results presented in \S\ref{sec:Results} will contain solutions that are the nonlinear analogue of a type $(1,n)$-wave.

\section{The numerical method} \label{sec:num}

\noindent  In this section, we describe the numerical procedure for solving Bernoulli's equation \eqref{eq:PhiBern} and the harmonic relationship \eqref{eq:Harm} for $X(\phi)$ and $Y(\phi)$ subject to a given value of the energy, $\E$, from \eqref{eq:Energy}. Thus: 
\begin{subequations}\label{eq:Main}
\begin{equation}
\label{eq:MainEq}
\frac{F^2}{2J}+Y+B\frac{(Y_{\phi}X_{\phi \phi}-X_{\phi}Y_{\phi \phi})}{J^{\frac{3}{2}}}=\frac{F^2}{2},
\end{equation}
\begin{equation}
\label{eq:MainHarm}
X_{\phi}(\phi)=1-\mathscr{H}[Y_{\phi}(\phi)],
\end{equation}
\begin{equation}\label{eq:MainEn}
  \E = \frac{F^2}{2 \Ehigh}\int_{-\frac{1}{2}}^{\frac{1}{2}}Y(X_{\phi}-1) \de \phi +\frac{B}{\Ehigh} \int_{-\frac{1}{2}}^{\frac{1}{2}}(\sqrt{J}-X_{\phi}) \de \phi +\frac{1}{2 \Ehigh}\int_{-\frac{1}{2}}^{\frac{1}{2}}Y^2 X_{\phi} \de \phi.
\end{equation}
\end{subequations}
Recall we define $E_\text{hw}$ as the energy of the highest Stokes wave ($E_\text{hw} \approx 0.00184$).

Solutions of the above problem are regarded as lying within $(B, F, \E)$-space. We solve these equations using Newton iteration on a truncated Fourier series using the fast Fourier transform. The procedure is as follows. 
\begin{enumerate}[label=(\roman*),leftmargin=*, align = left, labelsep=\parindent, topsep=3pt, itemsep=2pt,itemindent=0pt ]
\item An initial guess for $Y(\phi)$ is carefully chosen using either linear theory \eqref{eq:linearXY} or from a previously-computed solution [cf. \S\ref{sec:Results} for specific details].

\item Of the triplet $(B, F,\E)$, we choose to fix two parameters and treat the last parameter as an unknown.
%
\item The collocation variable, $\phi$, is discretised using $N$ grid-points, with $\phi_k = -1/2 + k \Delta \phi$ for $k = 0, \ldots, N-1$ and $\Delta \phi = 1/N$. We define the solution $Y(\phi_k) = Y_k$ at each of these points. Note that once $Y_k$ is known for all $k$, $X_k$ can be calculated using the harmonic relation \eqref{eq:MainHarm}. 

\item Combined with the unknown parameter (either $B$, $F$, or $\E$), this yields $N+1$ unknowns. Bernoulli's equation \eqref{eq:MainEq}, evaluated at $\phi_k$, provides $N$ equations and the system is closed with the additional energy constraint \eqref{eq:MainEn}. Newton iteration is then used to solve the nonlinear system of equations until a certain tolerance (typically $10^{-11}$) on the norm of the residual is met.
\end{enumerate}
\noindent In our numerical scheme, we leverage the Fourier Transform for efficient manipulation of the solutions. 
In particular, note that the Hilbert transform, $\mathscr{H}$, needed for the harmonic relation \eqref{eq:MainHarm}, can be evaluated via $\mathscr{H}[Y]=\mathscr{F}^{-1}[\i \cdot \text{sgn}(k)\mathscr{F}[Y]]$, where $\mathscr{F}$ denotes the Fourier transform and sgn is the signum function. 
Both the Fourier and inverse-Fourier transforms are calculated with the FFT algorithm. 
The derivatives of $Y$ are also computed in Fourier space using the relationship $Y^{(n)}(\phi)=\mathscr{F}^{-1}[(2 \pi \i k)^n \mathscr{F}[Y]]$. 
In order to obtain the numerical results presented in \S\ref{sec:Results}, we find it sufficient to use $N = 1024$ mesh points. 

In essence, our goal will be to study the $(B, F, \E)$-solution space, particularly as $B \to 0$. 
We start from a low-energy solution, and increase the parameter $\E$ until the desired value is reached. 
In order to initialise this continuation procedure at small values of $\E$, we select an initial Bond number which is chosen away from the Wilton-ripples value of $B_\text{wilton}$ in \eqref{eq:linearwiltonBF}. 
Then the Froude number approximated by the linear dispersion relation \eqref{eq:disp} with $k = 1$, and we use the linear approximations of $X$ and $Y$ from \eqref{eq:linearXY} with a small arbitrary choice of $\ep a_k$ (typically $10^{-5}$). 
For this linear solution, $\E$ is then calculated; the above serves as the initialisation procedure for the Newton scheme which solves for values of $Y_i$ and $\F$. 

Once solutions are found at desired values of $\E$, we establish the $(B, F)$-bifurcation space by continuation from a previously calculated solution. 
Note that in some cases, it will be necessary to fix $B$ or $F$ and solve for the other value, depending on the gradient of the bifurcation curves. 

\section{Numerical results for fixed energy, $\E$} \label{sec:Results}

\begin{figure}\centering
\includegraphics[width=0.4\textwidth]{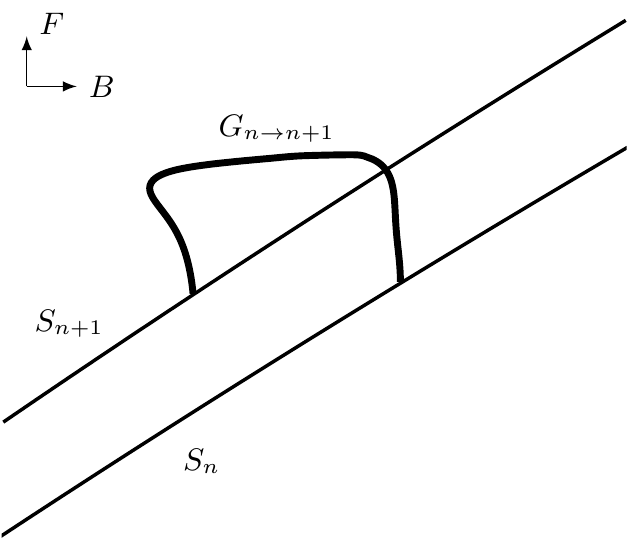}
\caption{A typical component of the bifurcation diagram illustrated in $(B, F)$-space consisting of a single finger, $G_{n \to n+1}$, (shown bolded) and two side curves $S_n$ and $S_{n+1}$. 
This is considered at a fixed value of the energy, $\E$. 
 \label{fig:roughsketch}}
\end{figure}

\noindent The numerical results we now present suggest that at a fixed value of $\E$, certain solutions in the $(B, F)$-bifurcation space can be classified according to `finger'-type structures and `side-branch'-type structures. An example of this structure, as drawn in the $(B,F)$ plane, is shown in fig.~\ref{fig:roughsketch}.

First, let us first define the \emph{side branch}, $S_n$, as 
\[
S_n = \{ \text{Bifurcation curve of solutions analogous to type $(0, n)$-waves}\}.
\]
Thus $S_n$ corresponds to those points in $(B,F)$-space associated with a certain type of solution. These solutions are pure $n$-waves ($1/n$-periodic solutions in the interval); they are the nonlinear analogue of the linear type $(0, n)$-waves introduced in \S\ref{sec:Linear}, \emph{i.e.} a sine or cosine wave with wavenumber $n$ about a constant mean value. 

In addition, adjacent side branches are connected by \emph{fingers}, say $G_{n \to n+1}$. We define such a structure as 
\[
G_{n \to n+1} = \{ \text{Bifurcation curve of solutions connecting $S_{n}$ to $S_{n+1}$} \}. 
\]
The finger can be interpreted as follows. Along $S_n$, solutions are pure $(n)$-waves; following this set of solutions, there exists a bifurcation point where the $1$-mode grows. Following this new branch, which is labeled $G_{n \to n+1}$, yields a solution analogous to a type $(1, n)$-wave. Continuing along $G_{n \to n+1}$, the solution transitions to type $(1, n+1)$ and then finally to pure-$(n+1)$ wave where it connects to $S_{n+1}$. 
An illustration of these classifications is shown in fig.~\ref{fig:roughsketch}.

In the following sections we present solutions along the side branches, $S_n$, and fingers, $G_{n \to n+1}$, for waves that are approximately half the height of the highest fundamental gravity wave.
For our choice of energy in \eqref{eq:MainEn}, this occurs at $\E=0.3804$.
Starting in \S\ref{sec:finger} we describe the structure of solutions across a prototypical finger, $G_{13 \to 14}$, and then in \S\ref{sec:sidebranch} we demonstrate how this finger bifurcates from side branches $S_{13}$ and $S_{14}$. 
Multiple fingers are then shown in \S\ref{sec:Btozero} for $n=7$ to $28$, demonstrating their behaviour as $B \to 0$.

\subsection{Analysis of a single finger, $G_{n \to n+1}$} \label{sec:finger}

\noindent  The prototypical finger $G_{13 \to 14}$ is shown in fig.~\ref{fig:finger1} for a value of $\E = 0.3804$. Note that solutions near the `tip' of the finger seem to correspond to the phenomena of parasitic ripples discussed in \S\ref{sec:Intro}; that is, we observe a series of small-scale capillary-dominated ripples riding on the surface of a steep gravity wave. This is shown in insets $(c)$, $(d)$, and $(e)$ in fig.~\ref{fig:finger1}. Below, we will continue to refer to solutions as being separated into capillary ripples and an underlying gravity wave, even though this classification may be ambiguous. 

As we move down either side of $G_{13 \to 14}$ by decreasing the Froude number, the amplitude of the ripples increases while the amplitude of the underlying gravity wave decreases. This is shown in fig.~\ref{fig:finger1} via the transitions $(c)\to (b)\to (a)$ and $(e)\to (f)\to (g)$. It becomes extremely challenging to numerically compute solutions below $(a)$ and $(g)$. 

Finally, as we travel from right to left across the finger, the wavelength of the ripples decreases as an extra ripple is formed. This can be seen by comparing solutions in insets $(g)$ and $(a)$ where $(g)$ has 13 maxima and $(a)$ has 14 maxima. The increase in the number of ripples can be observed as occurring near the tip of the finger between insets $(c)$ and $(e)$. We will discuss the structure of this process in \S\ref{sec:Discussion}. 

\begin{landscape}
\thispagestyle{lscape}
\pagestyle{lscape}
\begin{figure}
\includegraphics[scale=1, angle=90]{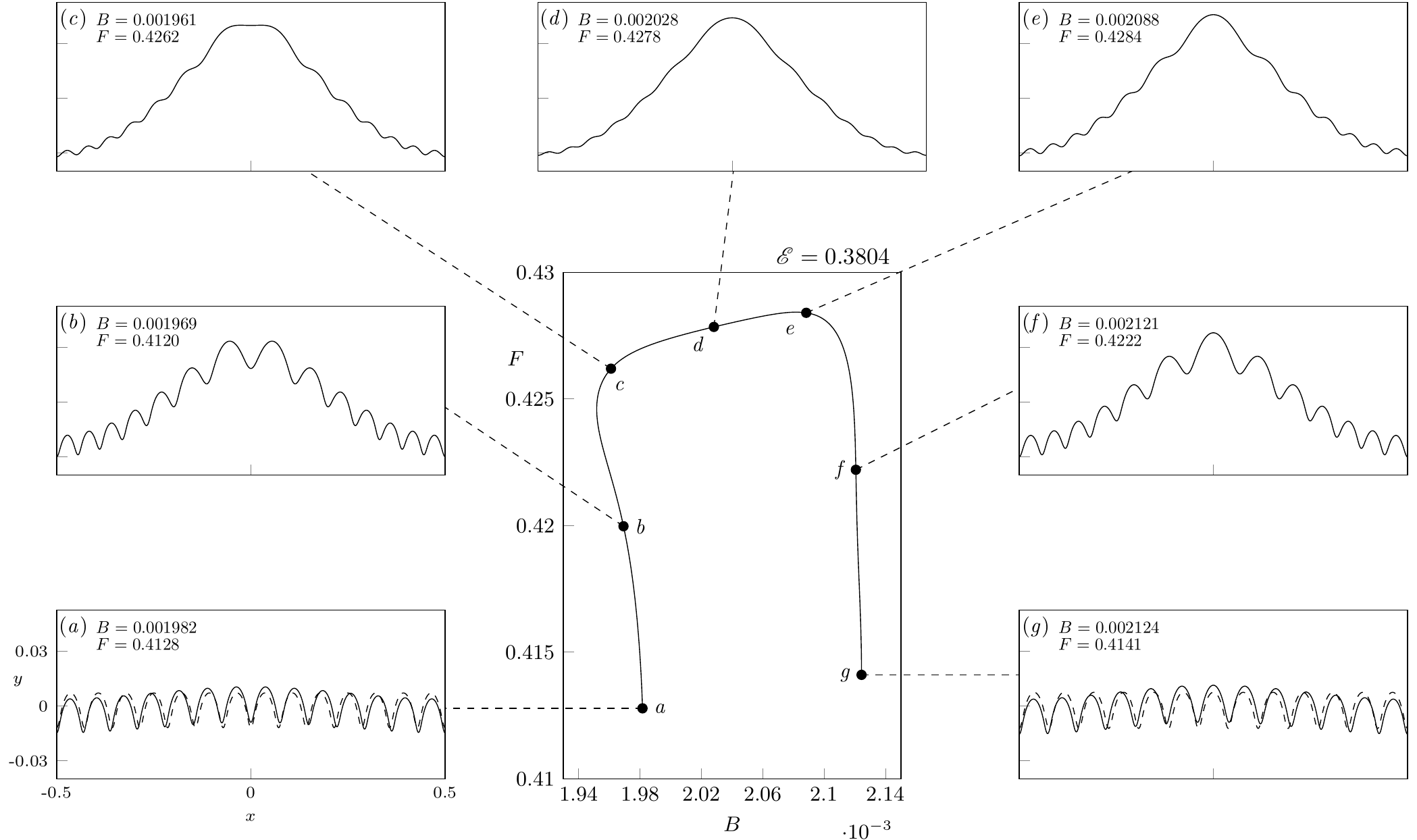}
\caption{\label{fig:finger1} A single finger of solutions, $G_{13 \to 14}$, is displayed in the $(B,F)$ plane for a value of $\E=0.3804$. The dashed profiles in $(a)$ and $(g)$ are solutions from branches $S_{13}$ and $S_{14}$ near the bifurcation point.}
\end{figure}
\end{landscape}

\subsection{Analysis of side branches $S_n$ and $S_{n+1}$} \label{sec:sidebranch}

\begin{figure}\centering
\includegraphics[scale=1]{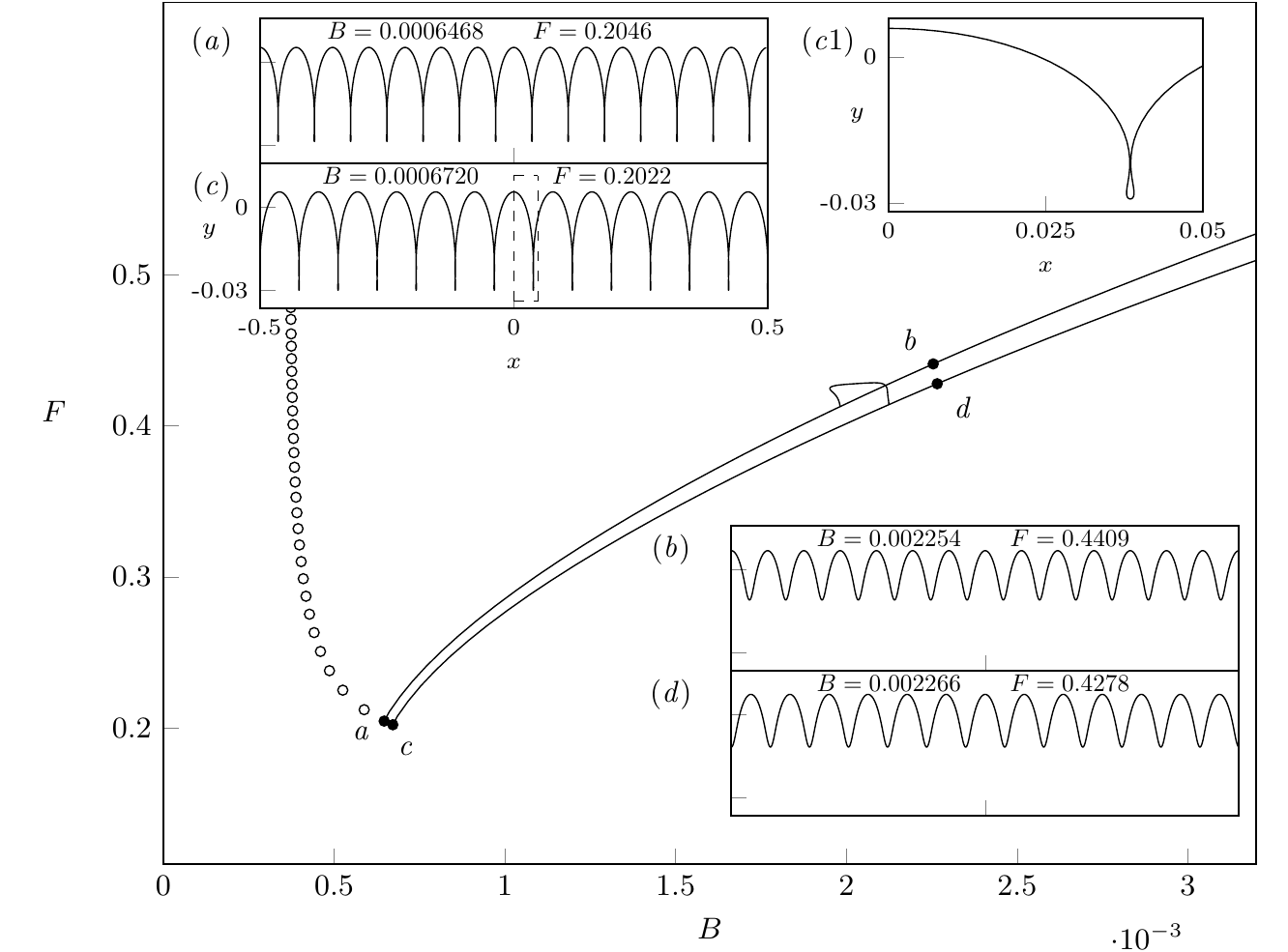}
\captionof{figure}{\label{fig:finger2} The finger $G_{13 \to 14}$ is shown against the two side branches $S_{13}$ and $S_{14}$. The two side branches terminate at points $a$ and $c$ (black circle) through the trapping of bubbles. Circles represent the locations where solutions of $S_{n}$ for $n \geq 15$, shown for every 5th point, become unphysical, found from the numerical predictions of appendix \ref{sec:NewSol}.}
\end{figure}

\noindent  We now discuss the side branches. In the case of the prototypical finger, $G_{13 \to 14}$, displayed in figure~\ref{fig:finger1}, we observe that this finger connects two side branches $S_{13}$ and $S_{14}$, as shown in figure~\ref{fig:finger2}. The branch $S_{13}$ contains pure $(13)$-waves, which have a fundamental wavelength of $\lambda=1/{13}$. The branch $S_{14}$ consists of pure $14$-solutions, which have $\lambda=1/{14}$.


We next observe that at fixed energy, $\E=0.3804$, the solutions in $S_{13}$ and $S_{14}$ reach a limiting configuration through the trapping of bubbles, shown by solutions $(a)$ and $(c)$. These branches of solutions are the large amplitude analogue of those predicted by linear theory in \S\ref{sec:Linear}, given by
\begin{equation}
F^2 \sim {1/(2n\pi)+(2n\pi)B} \qquad \text{and} \qquad 
F^2 \sim {1/(2\pi(n+1))+2\pi(n+1)B}.
\end{equation}
These were obtained by taking the values of $k=n$ and $k=n+1$ in the linear dispersion relation \eqref{eq:disp}.

In order to compute these branches numerically, an initial pure--$n$ solution was taken from linear theory with the dispersion relation \eqref{eq:disp} satisfied for $k=n$. 
This gives a cosine profile with $n$ peaks across the periodic domain. 
Slowly increasing the energy of this solution across multiple runs yields a single solution for each branch at $\E=0.3804$, from which these branches were calculated by continuation at fixed $\E$.

The location along the branch for which solutions reach a limiting configuration through a trapped bubble can be numerically predicted by the results of Appendix \ref{sec:NewSol}. 
These points are shown in figure~\ref{fig:finger2} for $n \geq 15$.

\noindent We see that as the value of $n$ for these limiting solutions increases, the value of $F$ at these points increases beyond that of the original finger.
Thus, below a certain value of the Bond number, we expect that each finger will instead bifurcate from nonphysical solutions. 
As we then proceed to increase the Froude number and transverse the side of each of these fingers for $B<B_{\text{crit}}$, we anticipate that the solutions will turn physical. 
This would result in the tip of each finger consisting of purely physical solutions.

\subsection{The unveiled structure for $B \to 0$} \label{sec:Btozero}


\noindent  This process of generating an individual finger may be repeated across different values of the Bond number, resulting in a remarkable structure that holds in the limit of $B \to 0$. 
Many of these fingers are shown in figure~\ref{fig:finger3} below, from $n=7$ to $n=28$; for clarity the side branches have been omitted from this figure.
As the Bond number decreases over each finger, the wavelength of the ripples decreases from $1/n$ to $1/(n+1)$, resulting in the formation of an additional crest. Consecutive fingers are connected at the point from which they bifurcate from the side branches of pure $n$-waves, demonstrated previously in \S\ref{sec:sidebranch} and shown by solutions $(d1)$ and $(d2)$ in figure~\ref{fig:finger3}. The solutions at these bifurcation points display a phase shift of $1/n$ between them. Due to this phase shift, the $n$th Fourier coefficient changes sign between these solutions at this bifurcation point. It is this phase shift that led \cite{chen1979steady} to misleadingly state (on p.~204) that the weakly nonlinear solutions are discontinuous with respect to the $n$th Fourier coefficient at this point.

From solutions $(a1)$, $(b)$ and $(c)$, labeled at the top of the fingers in figure~\ref{fig:finger3}, we observe that as $B \to 0$, the amplitude of the ripples decreases and the overall solution appears to tend towards the fundamental Stokes wave with energy $\E$. Although the profile in subfigure~\ref{fig:finger3}$(a1)$ seems to indicate a pure gravity wave, the capillary ripples can be detected under closer inspection. In order to quantify this, we isolate the pure gravity wave solution, $y_0$, and plot $y - y_0$ in subfigure~\ref{fig:finger3}$(a2)$. This shows that the ripples are still present in the solution, but with a very small amplitude. Moreover, it can be verified that the profile norm, $\lvert y-y_0 \rvert = O(B)$, by repeating this procedure for multiple solutions along the top of the fingers in figure~\ref{fig:finger3}. We shall comment on this algebraic error and the exponentially-small ripples in \S\ref{sec:Discussion}. 

We note that the presence of this bifurcation from the side branches $S_n$ to the fingers $G_{n \to n+1}$ can be observed by a change in sign of the Jacobian along $S_n$ as the bifurcation point is passed. Solutions close to this bifurcation point are shown by $(a)$ and $(g)$ in figure \ref{fig:finger1} for $S_n$ (dashed) and $G_{n \to n+1}$ (solid).

Furthermore, the range of $F$ between the tip of each finger and the bottom remains of $O(1)$ as $B \to 0$ for the solutions calculated in figure~\ref{fig:finger3}. 
Consider for instance the range between solutions $(a1)$ and $(f)$.
This suggests the existence of an interval of solutions holding under the $B \to 0$ limit. 
The solution with the largest value of $F$ is expected to

\begin{landscape}
\thispagestyle{lscape}
\pagestyle{lscape}
\begin{figure}
\includegraphics[scale=1, angle=90]{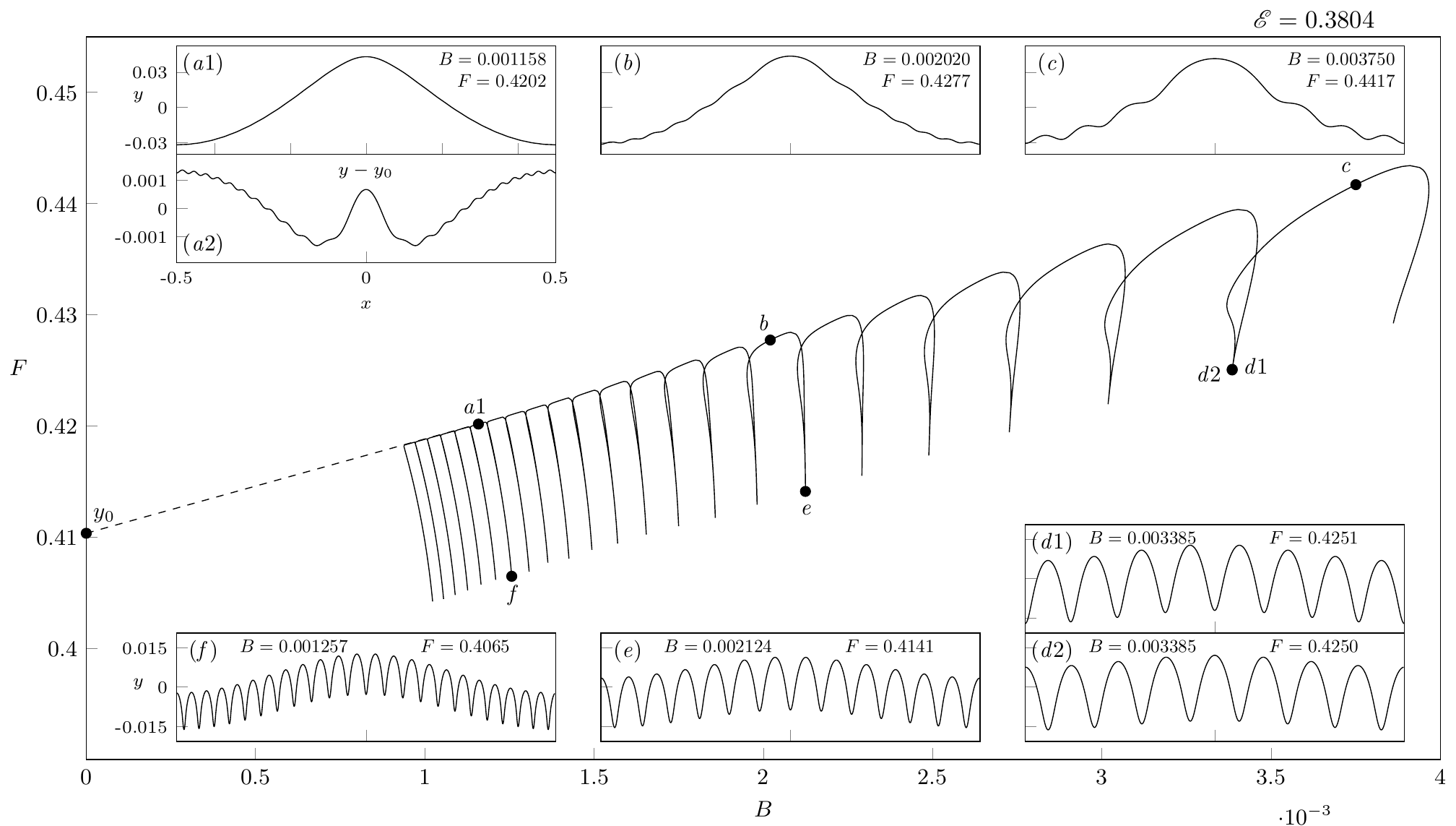}
\caption{\label{fig:finger3} Multiple fingers $G_{7 \to 8}$ to $G_{28 \to 29}$ are shown to form one connected branch in the $(B,F)$ plane at fixed $\E=0.3804$.}
\end{figure}
\end{landscape}

\noindent be the fundamental Stokes wave with $B=0$ and $\E=0.3804$, shown by the point $y_0$ in figure~\ref{fig:finger3}. 
We predict that, as $B \to 0$, the solutions with smallest Froude number in this interval are unphysical due to an intersecting free-surface. 
This is because, for $B<B_{\text{crit}}$, the pure $n$-solutions near the bifurcation point on the side branches are anticipated to also be unphysical.
The interval would then contain a range of solutions, which transition from unphysical to physical as the Froude number increases.

\begin{figure}\centering
\includegraphics[scale=1]{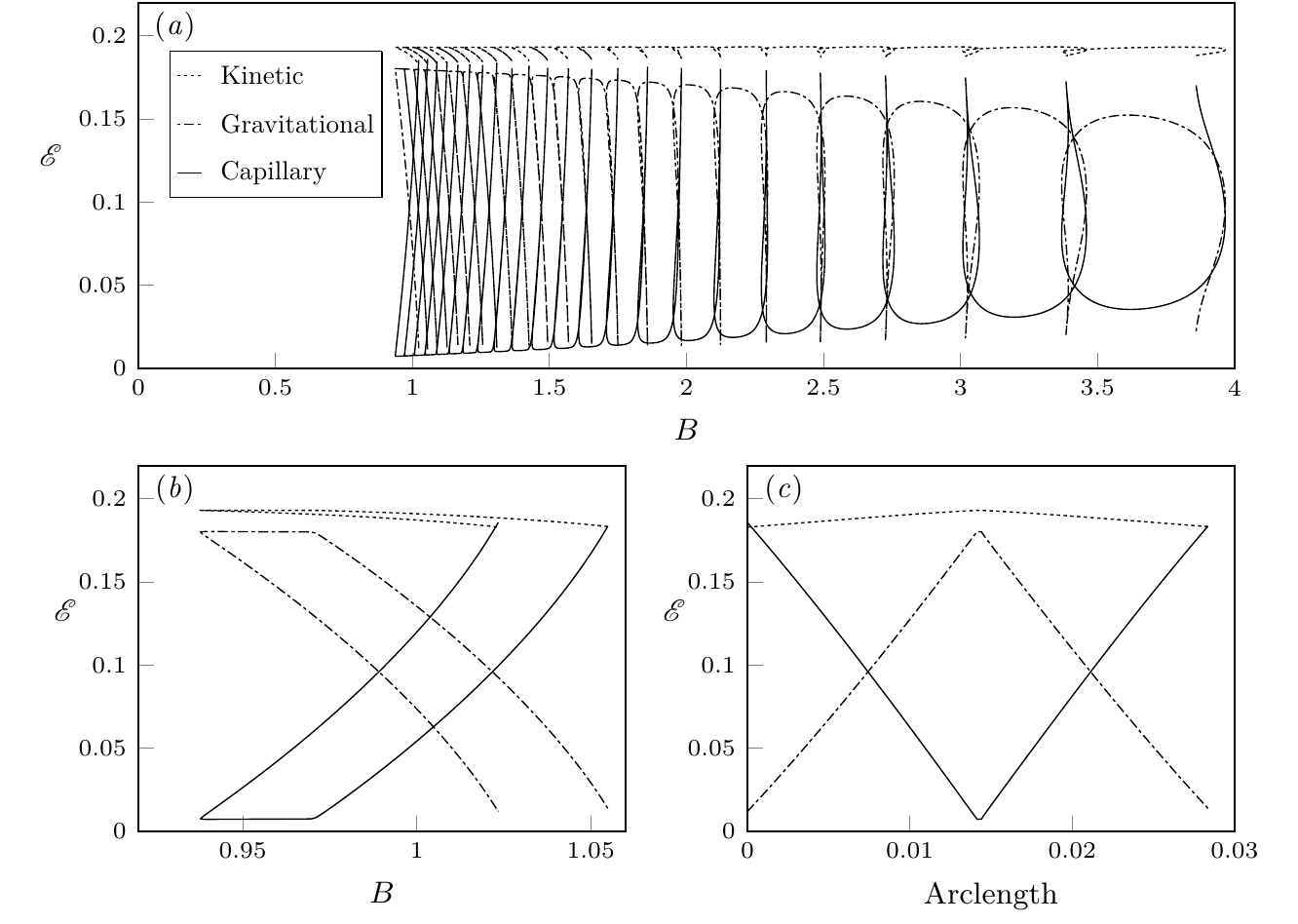}
\captionof{figure}{\label{fig:energytran} (a) The corresponding kinetic (dotted), gravitational (dash-dotted), and capillary (solid) energies for the solutions from figure \ref{fig:finger3}. In the two lower figures, the kinetic, capillary, and gravitational energies are also shown for (b) $B$ vs. $\E$ and (c) branch arclength vs. $\E$ for the solutions corrresponding to the single finger $G_{28 \to 29}$.}
\end{figure}

\section{Relation to previous numerical attempts}
\label{sec:Pastwork}
%

\noindent A key challenge is to understand the relationship between our solutions of fixed energy, $\E$, and those of previous authors with a different amplitude condition, say $\mathscr{A}$. In this section we demonstrate that a key limitation of previous choices of amplitude is the existence of highly energetic (and subsequently nonlinear) solutions at small values of $\mathscr{A}$. Thus, somewhat surprisingly, alternative choices of the amplitude measure may admit nonlinear solutions in the naive linear limit of $\mathscr{A} \to 0$ -- this occurs due to the singular nature of $B \to 0$ and in particular, the nature of the solutions between adjacent fingers.

\subsection{Solutions at different values of the energy $\E$} \label{sec:VaryingE}

\noindent  In the previous section, we demonstrated the structure of the bifurcation diagram and associated solutions at fixed energy $\E=0.3804$. In fact, this bifurcation structure is only perturbed in a regular fashion as the energy changes near this value. Thus, the full structure of solutions, which holds as $B \to 0$, can be computed for different values of $\E$ in a straightforward manner. 

\begin{figure}
\begin{centering}
\includegraphics[scale=1]{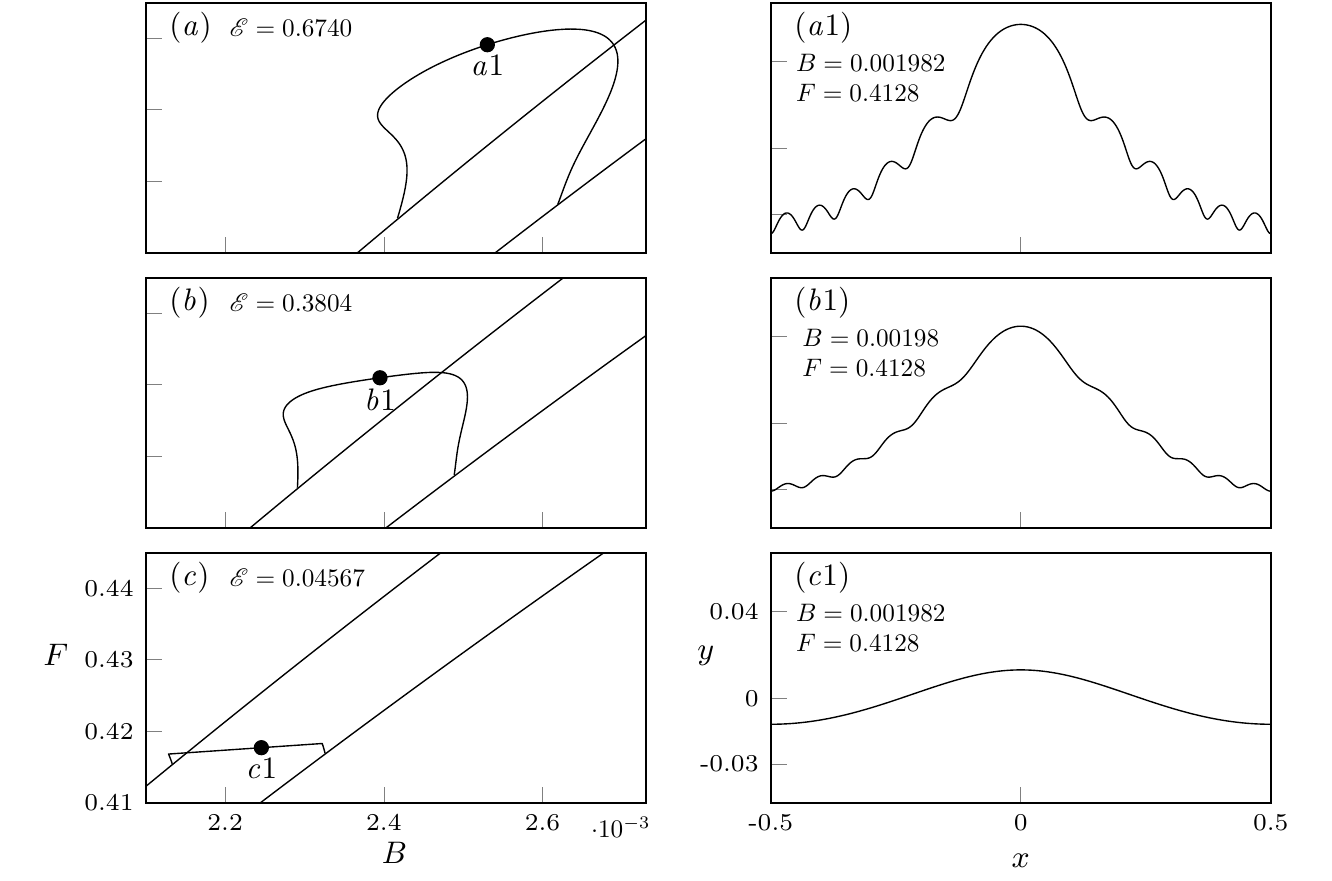}
\end{centering}
\caption{\label{fig:energylevels} Shown in the left subplots are the fingers $G_{11 \to 12}$ and the side branches $S_{11}$ and $S_{12}$, plotted in the $(B,F)$-plane, for $(a)$ $\E = 0.67$; $(b)$ $\E = 0.3804$; $(c)$ $\E = 0.046$. Example solutions, near the tops of the fingers, are shown in the corresponding right subplots labeled $(a1)$, $(b1)$, and $(c1)$.}
\end{figure}

We show an example of this in figure~\ref{fig:energylevels}, where we display the finger $G_{11 \to 12}$ and the side branches $S_{11}$ and $S_{12}$ for three different values of $\E$. In the figure, the value of $\E$ decreases from $\E=0.67$ in $(a)$ to $\E=0.3804$ in $(b)$ to $\E=0.046$ in $(c)$. Three changes to either the solution or branch structure are noticeable as $\E$ decreases:
\begin{enumerate}[label=(\roman*),leftmargin=*, align = left, labelsep=\parindent, topsep=3pt, itemsep=2pt,itemindent=0pt ]
  \item The amplitude of the ripples decreases.
  \item The range of $F$ between the top and bottom of the finger decreases.
 \item  The finger becomes more rectangular.
\end{enumerate}
In $(c)$, the amplitude of the ripples has decreased to the point at which they are no longer observable visually.

\subsection{Choice of amplitude parameters in previous works} \label{sec:PastworkAmp}

\noindent  We now re-visit the alternative choices of the amplitude or energy parameter in \eqref{eq:Amplitudes}. A few of the solutions displayed in figure~\ref{fig:finger3} are similar to those previously calculated by \cite{schwartz1979numerical}, who plotted remnants of this figure at larger values of $B$ for a different amplitude parameter. Since their choice of amplitude,
\begin{equation} \label{eq:VB}
\mathscr{A} = A \equiv [y(0)-y(\pi)]/2\pi,
\end{equation}
relies on local values at the centre and edge of the periodic domain, they found these branches to behave somewhat differently than how we have described them in our \S\ref{sec:Results}. 

Notice that according to their choice of norm \eqref{eq:VB}, waves with an even number of crests that are equally spaced throughout the domain will have $y(0)=y(\pi)$ and consequently $A=0$. This corresponds to every other branch of solutions with fundamental wavelength smaller than the periodic domain, $S_{n}$ with $n$ even. 
Hence for the branch of solutions $G_{n \to n+1}$ at fixed $\E$, $A$ grows smaller tending towards solutions near the bifurcation points of $S_{n+1}$ and $S_{n}$---despite the high-nonlinearity of these solutions.
This is demonstrated for $n=13$ in figure~\ref{fig:finger1} with solution $(a)$, which approaches $S_{14}$, and solution $(g)$, which approaches $S_{13}$.
Thus the bifurcation from $S_{14}$ connecting finger $G_{14 \to 15}$ to $G_{13 \to 14}$ will occur from an amplitude value of $A=0$. This is one reason why the full structure of solutions was not revealed through smooth continuation at fixed $A$ by the investigations of \cite{schwartz1979numerical}.

Next, let us turn to the numerical investigation of the $B \to 0$ limit performed by \cite{chen1980steady}, who fixed the first Fourier coefficient,
\begin{equation*}
\label{eq:CS}
\mathscr{A}=A_1,
\end{equation*}
 as an amplitude parameter. We now know that, since the bifurcation between distinct fingers in the $(B,F)$ plane occurs via the side branches $S_n$, which have a first Fourier coefficient of zero for $n \geq 2$, it is impossible to recover the structure shown in figure~\ref{fig:finger3} with a fixed value of $A_1$. \cite{chen1980steady} had indicated the impossibility of a continuous deformation to the pure Stokes gravity wave as $B \to 0$, but we now see that this occurred as a direct consequence of their chosen amplitude parameter.  

\subsection{An insufficient number of Fourier coefficients} \label{sec:PastworkFourier}

\noindent  As we have noted, it is crucial to select the right continuation parameter in order to recover the $B \to 0$ limit. There are other possible reasons why others may have struggled to reproduce an accurate structure of the parasitic ripple phenomena. In particular, a large number of Fourier modes are required in order to capture the regions between adjacent fingers, and this is primarily due to the bifurcation occurring from the side branches, $S_{n}$, which contains solutions that approach pure $n$-waves. Thus, solutions within the finger $G_{n \to n+1}$, which are located near to side branches are then dominated by the $n$th Fourier coefficient. If in our numerical scheme we consider a series truncation at the $N$th Fourier coefficient, then the main coefficients contributing to the capillary-dominated ripples will be a multiple of $n$. Hence, an effective number of $N/n$ Fourier coefficients will describe the behaviour of the wave near to this bifurcation point.

For the computation of the gravity-capillary wave with the parasitic ripples (their fig.~10) \cite{schwartz1979numerical} used $N = 40$ in order to capture a wave with $n=11$ ripples. Thus, in order to investigate the side-branch bifurcation associated with this solution, their Fourier expansions would have contained an effective number of $N/n \approx 4$ Fourier coefficients---which is insufficient.
Within this work, we have been using $1024$ Fourier coefficients, which corresponds to $35$ effective coefficients for solutions near the bifurcation point of the finger with smallest Bond number in figure~\ref{fig:finger3}.



\subsection{The symmetry shifting bifurcation}
\label{sec:PastworkSym}

\begin{figure}
\begin{centering}
\includegraphics[scale=1]{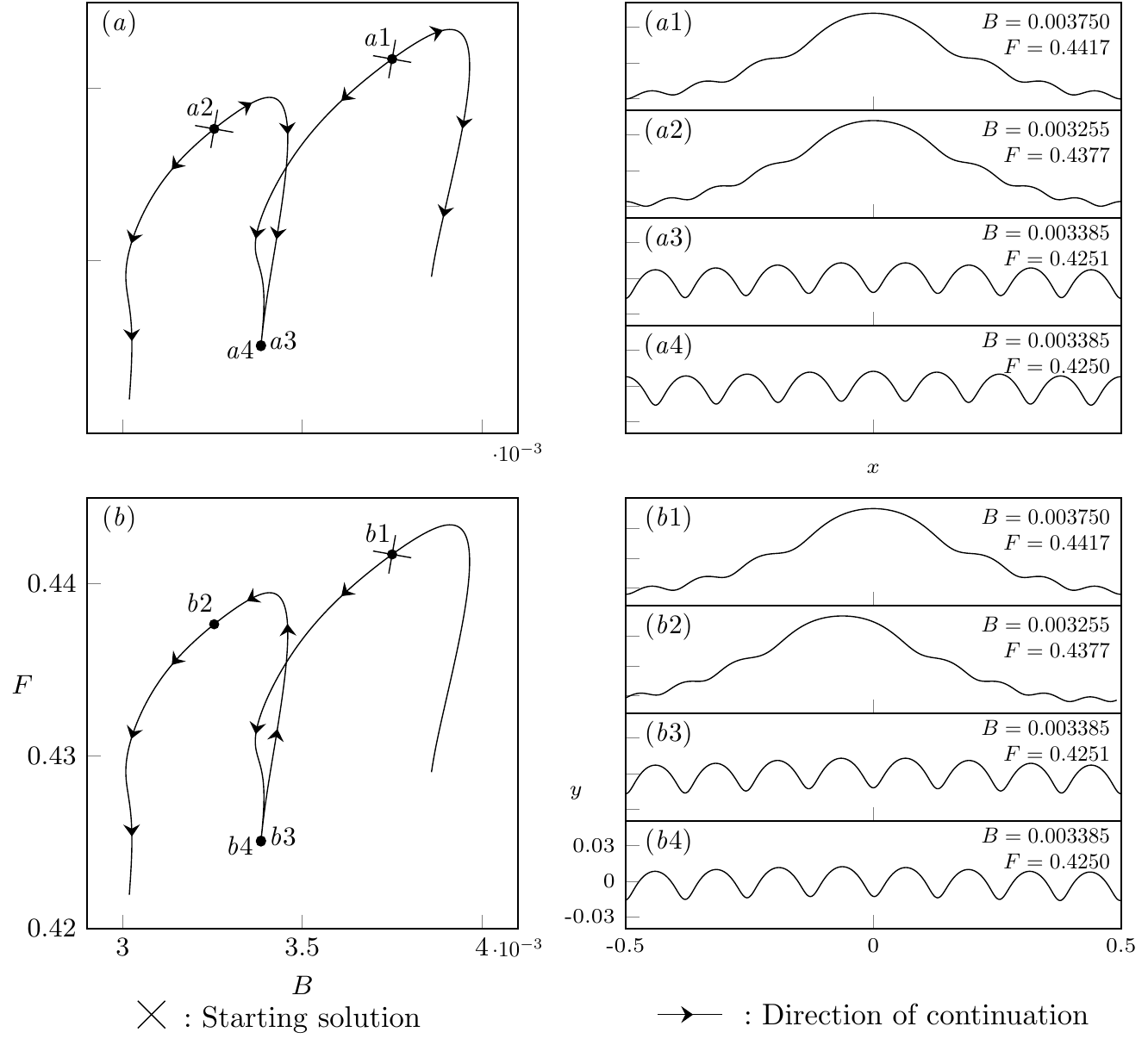}
\end{centering}
\caption{\label{fig:phase1} Two methods for numerical continuation are depicted. The starting location for continuation is denoted by a cross, and the arrows indicate the direction travelled by continuation.}
\end{figure}

In addition to the importance of selecting the correct amplitude measure, let us discuss the relationship between the bifurcation structure presented earlier (e.g. in our figure~\ref{fig:finger3}) with the constraint on the symmetry in the travelling-wave frame. For the solutions displayed in figure~\ref{fig:finger3}, each finger is computed beginning with an initial solution that lies on the finger and then, with a fixed $\E$, solutions are obtained by continuation to either side of the starting point until the entire finger is computed. As a result of this continuation scheme, the solutions at the bottom of adjacent fingers are out of phase with one another; this can be seen in solutions $(d1)$ and $(d2)$ in figure~\ref{fig:finger3}. 
This method of continuation is depicted more clearly in figure~\ref{fig:phase1}$(a)$, where $(a1)$ and $(a2)$ are two starting solutions, while $(a3)$ and $(a4)$ are out of phase. 

\noindent  Finally, let us discuss the relationship between the bifurcation structure presented earlier with the imposition of symmetry in the travelling-wave problem. For the solutions displayed in figure~\ref{fig:finger3}, each finger is computed beginning with an initial solution that lies on the finger and then, with a fixed $\E$, solutions are obtained by continuation to either side of the starting point until the entire finger is computed. As a result of this continuation scheme, the solutions at the bottom of adjacent fingers are out of phase with one another; this can be seen in solutions $(d1)$ and $(d2)$ in figure~\ref{fig:finger3}. 
This method of continuation is depicted more clearly in figure~\ref{fig:phase1}$(a)$, where $(a1)$ and $(a2)$ are two starting solutions, while $(a3)$ and $(a4)$ are out of phase. 

Alternatively, we could formulate a continuation scheme where the solutions in each finger are connected to those in adjacent fingers in a continuous fashion, depicted in figure~\ref{fig:phase1}$(b)$. Thus for example, the scheme is started with a single initial point, $(b1)$, shown in figure~\ref{fig:phase1}. This finger is then found via the typical continuation method. Having located a solution, $(b3)$, at the bifurcation point, the adjacent finger is completed by using $(b3)$ as a starting solution for continuation.
This alternative method shown in figure~\ref{fig:phase1}$(b)$ results in solutions $(b3)$ and $(b4)$ at the bottom of consecutive fingers with no phase shift. The result of this approach is a continuous set of solutions as $B \to 0$. 

In using this alternative method, solutions at the top of consecutive fingers have a shifted point of symmetry, as demonstrated by comparing solutions $(b1)$ and $(b2)$ in figure~\ref{fig:phase1}. 
This point of symmetry has been moved from $x=0$ to $x=-1/n$ for all solutions on the new finger. 
We denote this to be a \emph{symmetry shifting bifurcation}, which is unable to be captured if the point of symmetry of the wave is prespecified. This assumption of a fixed point of symmetry is often used in the two following methods:
\begin{enumerate}[label=(\roman*),leftmargin=*, align = left, labelsep=\parindent, topsep=3pt, itemsep=2pt,itemindent=0pt ]
  \item Numerical procedures which solve for the half-domain $x=[0,1/2]$ and enforce a turning point at $x=0$, such as that by \cite{schwartz1979numerical}.
  \item  The analytical work of \cite{chen1979steady}, who posit a weakly nonlinear solution with assumed symmetry at $x=0$.
\end{enumerate}
Both of these methods will be unable to capture this smooth $B \to 0$ limit with continuous solutions at the bifurcation point.

The relaxation of the fixed point of symmetry, in contrast to the assumption in \cite{chen1979steady}, is the modification required to correct their earlier statement on the validity of the $B \to 0$ limit for gravity capillary waves.


\section{Conclusions}
\label{sec:Conclusions}

\noindent  We have studied limitations in the previous works by \cite{schwartz1979numerical} and \cite{chen1979steady,chen1980steady}, highlighting three core issues. These are:
\begin{enumerate}[label=(\roman*),leftmargin=*, align = left, labelsep=\parindent, topsep=3pt, itemsep=2pt,itemindent=0pt ]
\item The chosen amplitude paramters, relying either on local values of the wave-height or specific Fourier coefficients (cf. \S\ref{sec:PastworkAmp}).

\item A small number of Fourier coefficients retained in the numerical schemes, the issues for which become more prominent near the bifurcation points (cf. \S\ref{sec:PastworkFourier}).

\item Assumptions made on the point of symmetry of the wave-profile (usually fixed to be at $x=0$), due to the symmetry-shifting bifurcation connecting adjacent fingers (cf. \S\ref{sec:PastworkSym}).
\end{enumerate}
In being aware of these, we have introduced alternative methods, either by solving or mitigating the issues. For instance we have used the wave-energy, $\E$, as an amplitude parameter. This has allowed us to be able to find a number of different types of solutions to the steep gravity-capillary wave problem existing under the limit of $B \to 0$. One of these, the steady symmetric parasitic ripple, is similar to the asymmetric parasitic waves encountered physically and will be the focus of our forthcoming analytical work.

\section{Discussion} \label{sec:Discussion}

\noindent In \S\ref{sec:Btozero}, we described the two types of solutions that are found as $B \to 0$. The first of these are found along the sides of the fingers and the side branches; they can be described by a multiple-scales type expansion that captures the rapid oscillations about a slowly varying mean. As they become increasingly oscillatory (with diminishing $F$), they reach an unphysical configuration through the trapping of bubbles for $B<B_{\text{crit}}$ (cf. end of \S\ref{sec:sidebranch}). Our focus has been moreso on the second of these types of solutions---those which correspond to waves with parasitic ripples lying on a gravity wave; these solutions are found near the tops of each finger. In a forthcoming work \citep{GCripples} we shall present an asymptotic theory for the description of these parasitic ripples. 

\begin{figure}\centering
\includegraphics[scale=1.25]{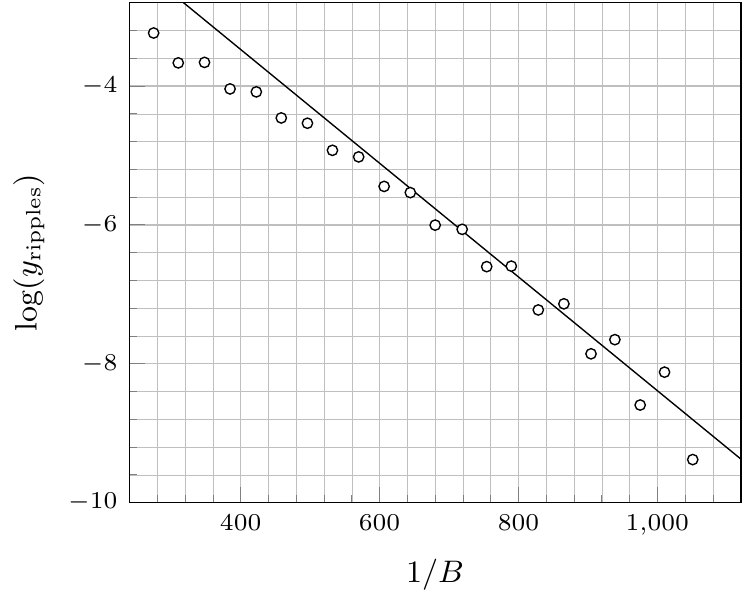}
\caption{\label{fig:expscale} The exponential scaling in equation \eqref{eq:expansion2} is shown (circles) for our numerical solutions by plotting $1/B$ v.s. $\log(y_{\text{ripples}})$. One numerical solution is chosen from each finger, corresponding to that of minimal ripple amplitude. From our forthcoming work, the analytical prediction (line), which depends on the value of $\E$, predicts a gradient of approximately $-8.2 \cdot 10^{-3}$.}
\end{figure}

The essential details are as follows. For those solutions that correspond to parasitic ripples on gravity waves, we may expand their form as a naive expansion in powers of the Bond number:
\begin{equation} \label{eq:expansion}
y(x) = \sum_{n=0}^{\infty}B^n y_n(x),
\end{equation}
such that in the limit of $B \to 0$ we recover the pure gravity-wave, $y_0$. However, as it turns out, the magnitude of the short-wavelength parasitic ripples is exponentially small in the Bond number, $B$. Thus 
\begin{equation}
\label{eq:expansion2}
\lvert y_{\text{ripples}} \rvert \sim e^{-\frac{\text{const.}}{B}}. 
\end{equation}
The above can be validated based on our numerical computations in the following way. For each finger, $\G{n}{n+1}$, a solution profile is calculated at the tip of the finger (i.e. the vertex in the $(B,F)$-plane). At this point, the magnitude of the parasitic ripple is approximated by examining $y - y_0$ and measuring the crest-to-trough amplitude for the oscillation nearest to the edge of the domain, $x = 1/2$ (a typical profile is shown in $(a2)$ of figure~\ref{fig:finger3}). The result of this numerical experiment is shown in figure~\ref{fig:expscale}; indeed the ripple amplitude lies approximately on a straight line in the semilog plot, confirming the exponential smallness of the ripples as $B \to 0$.

Thus in light of \eqref{eq:expansion2} these parasitic ripples will fail to be captured by \eqref{eq:expansion} and must be found \emph{beyond-all-orders} of the naive asymptotic expansion. Consequently, the use of specialised tools in asymptotic analysis, known as \emph{exponential asymptotics}, are required [see e.g. \cite{chapman_1998_exponential_asymptotics,chapman_2006_exponential_asymptotics,trinh_2013_new_gravity-capillary}]. Here, the necessary theory for prediction of the parasitic ripples is analogous, in spirit, to theories for the prediction of generalised solitary waves \citep{boyd1998weakly}, but there are a number of additional challenges due to the more involved boundary-integral framework and the lack of a closed-form leading-order Stokes solution, $y_0$.

We remark, in addition, that the idea of exponentially small parasitic capillary ripples in the classic Stokes-wave problem is not a new one. Indeed, as we discussed in \S\ref{sec:L-H} \cite{longuet-higgins_1963_the_generation} had proposed an analytical methodology for the derivation of parasitic ripples (followed by a similar approach in \cite{longuet1995parasitic}). However, both of these approaches are \emph{ad-hoc} in nature, and as noted by \cite{perlin1993parasitic}, fail to predict the correct magnitude of the ripples. Other theories, such as the averaged Lagrangian methodology of \cite{crapper1970non}, exhibit similar difficulties in providing rigorous comparison to numerical results---the latter author notes that ``\ldots \emph{[the theory] is not very accurate, but at the timing of writing is probably the best available}'' (p.~154). Consequently, the point we emphasise is that the systematic $B \to 0$ results we have presented in this paper are crucial for validation of the small surface tension limit. The presentation of a complete exponential asymptotics treatment of $B \to 0$ and the importance of prior approaches in inspiring such a methodology will be the focus of our forthcoming work.

Finally, we have only found symmetric solutions of the parasitic-ripples problem in this work. For general values of $B$ and $F$, and not necessarily only for the regime of small $B$, we note that there are extensive efforts to search for steady asymmetric solutions.
See for instance the works on gravity-capillary waves by \cite{zufiria1987symmetry} for finite depth, \cite{shimizu2012appearance} for infinite depth, and \cite{zufiria1987non} for pure gravity on infinite depth. Indeed, many of the asymmetric profiles in the works of e.g. \cite{shimizu2012appearance} exhibit similarities to the profiles shown in our work. Thus, it seems likely that the bifurcation structures we have presented in this work form a subset of a much more complicated structure that includes the potential for asymmetry. It remains to be seen if this asymmetry of the steady system would account for that observed in the experimental results, or whether it is necessary to consider unsteady flows, such as the time-dependant Navier-Stokes formulation considered numerically by \cite{mui1995vortical} and \cite{hung2009formation}.

\appendix
\section{The wave energy} \label{sec:Energy}

\noindent The non-dimensionalised bulk energy in the physical domain is given by 
\begin{equation}
\label{eq:EN2}
\overline{E}=\frac{F^2}{2}\int_{-\frac{1}{2}}^{\frac{1}{2}}\int_{-\infty}^{\zeta}(\phi_{x}^2+\phi_{y}^2) \de{y} \de{x} + B \int_{-\frac{1}{2}}^{\frac{1}{2}}\big([1+\zeta_{x}^2]^{\frac{1}{2}}-1\big) \de{x} +\int_{-\frac{1}{2}}^{\frac{1}{2}}\int_{-\infty}^{\zeta} y \, \de{y} \de{x}.
\end{equation}
On the right hand-side, the three groups correspond to the kinetic, capillary potential, and gravitational potential energies. 

Note that due to our subflow (where $\phi_x \to -1$ as $y \to -\infty$), the first and third integrals on the right hand side of \eqref{eq:EN2} will be unbounded. 
We thus define the energy, $E$, to be the the difference between \eqref{eq:EN2} and ``no-flow", $\zeta=0$, energy, which yields a finite value. 
This is then transformed to act on the free-surface, $\psi=0$, only 
 by the method of \cite{longuet1989capillary}, with which we change variables from $(x,y)$ to find the wave-energy under the $(\phi,\psi)$ mapping,
\begin{equation} \label{eq:EN3}
E = \frac{F^2}{2}\int_{-\frac{1}{2}}^{\frac{1}{2}}Y(X_{\phi}-1) \, \de{\phi} +B \int_{-\frac{1}{2}}^{\frac{1}{2}}(\sqrt{J}-X_{\phi}) \, \de{\phi}  + \frac{1}{2}\int_{-\frac{1}{2}}^{\frac{1}{2}}Y^2 X_{\phi} \de{\phi}.
\end{equation}
With this choice of amplitude parameter, $E_{\text{hw}} \approx 0.00184$ corresponds to the fundamental Stokes wave of maximum height. We rescale the energy by this value to obtain the amplitude parameter, $\E$, used within this report, given by $\E = \frac{E}{E_{\text{hw}}}$, in equation \eqref{eq:MainEn}.

\section{Limiting solutions of smaller fundamental wavelength}
\label{sec:NewSol}

\begin{figure}
\begin{centering}
\includegraphics[scale=1]{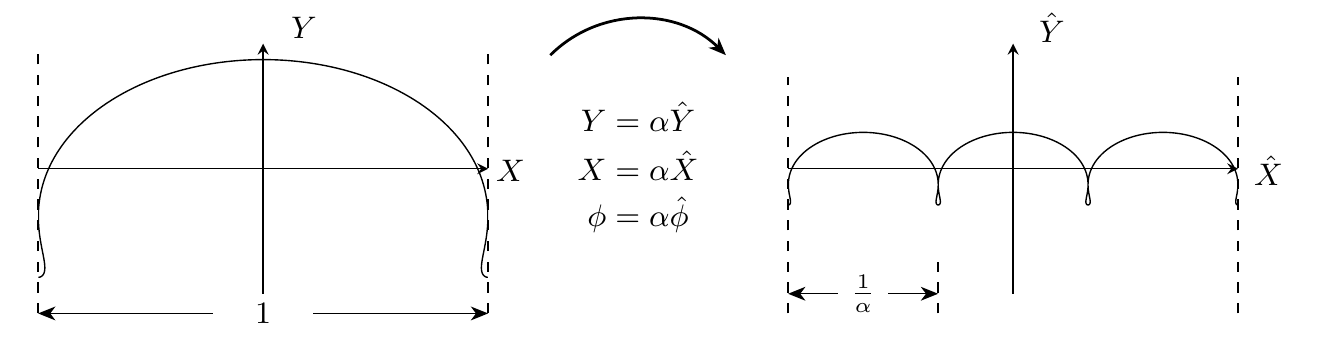}
\end{centering}
\caption{\label{fig:NewSol1} The rescaling used to produce another solution with smaller fundamental wavelength is shown.}
\end{figure}

\noindent If one solution is known to the gravity-capillary wave-problem with fundamental wavelength $\lambda=1$, another can be constructed with $\lambda=1/ \alpha$, where $\alpha$ is a positive integer.
This is visualised in figure~\ref{fig:NewSol1}.
Suppose we have a solution to Bernoulli's equation \eqref{eq:MainEq} and the harmonic relation \eqref{eq:MainHarm}.
In rescaling $Y=\alpha\hat{Y}$, $X=\alpha \hat{X}$, and $\phi=\alpha \hat{\phi}$, we repeat the first solution $\alpha$ times to map the original domain from $\phi \in [-\tfrac{\alpha}{2}, \tfrac{\alpha}{2})$ to the new domain $\hat{\phi} \in [-\tfrac{1}{2}, \tfrac{1}{2})$. 
This new solution, $\hat{X}$ and $\hat{Y}$, also satisfies the two governing equations with rescaled Froude and Bond numbers $\hat{F}$ and $\hat{B}$, given by
\begin{subequations}
\begin{equation}
\label{eq:NS1}
\hat{F}=\frac{F}{\sqrt{\alpha}} \quad \text{and} \quad  \hat{B}=\frac{B}{\alpha^2}.
\end{equation}
The energy of this new solution can be found by substituting the rescaled variables $\hat{X}$, $\hat{Y}$, $\hat{F}$, and $\hat{B}$ into equation \eqref{eq:MainEn}, yielding
\begin{equation}
\label{eq:NS2}
\hat{ \E}=\frac{ \E}{\alpha^2}.
\end{equation}
\end{subequations}
Since $\alpha>1$, the energy of the new pure $\alpha$-wave, $\hat{ \E},$ will always be smaller than that of the original wave, $\E$.

This ability to construct new solutions with fundamental wavelength shorter than the periodic domain allows us to numerically predict the point at which the side branches $S_{\alpha}$ transition from physical to unphysical solutions due to the trapping of bubbles.
These locations are shown in the bifurcation diagram of figure~\ref{fig:finger2}.
As all of these side branch solutions have the same energy, $\hat{\E}=0.3804$, but different values of $\alpha$, the energy of the original wave is given by $\E=\alpha^2 \hat{\E}$. 

The procedure to find the location at which solutions in $S_{\alpha}$ become unphysical is as follows:

\begin{enumerate}[label=(\roman*),leftmargin=*, align = left, labelsep=\parindent, topsep=3pt, itemsep=2pt,itemindent=0pt ]
 \item First, we numerically calculate the $(B,F)$ solution space of pure $1$-waves with a single trapped bubble amplitude condition. The energy of these solutions will vary.
  \item Second, we select the profile with $\E=\alpha^2 \hat{\E}$ and obtain values for $B$ and $F$.
  \item  Third, we rescale these by using equation \eqref{eq:NS1} to find $\hat{F}$ and $\hat{B}$. This yields the location at which solutions within the side branch, $S_{\alpha}$, become unphysical.
\end{enumerate}
Repeating this process for multiple values of $\alpha$ yields the predictions displayed with the circles in figure~\ref{fig:finger2}. With this method we are able to calculate solutions with a large value of $\alpha$ while keeping the number of Fourier coefficients used during Newton iteration fixed, and thus do not encounter the issue discussed in \S\ref{sec:PastworkFourier}.

\begin{figure}
\begin{centering}
\includegraphics[scale=1]{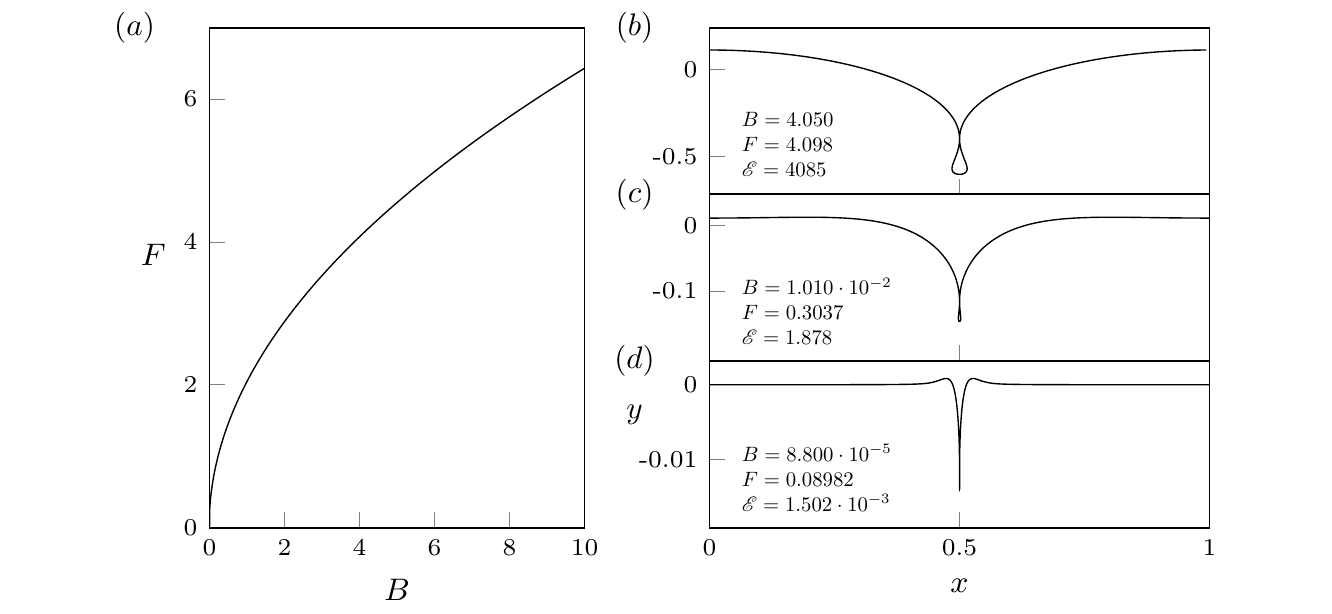}
\end{centering}
\caption{\label{fig:NewSol2} $(a)$ The branch of pure-$1$ solutions displaying an enclosed bubble is shown in the $(B,F)$ bifurcation diagram. Note that we have displayed the domain $x \in [0,1]$ to demonstrate this limiting behaviour. }
\end{figure}

It would also be possible to use this method to compute all of the solutions along the side branch $S_{\alpha}$. 
However, this requires the entire sheet of pure $1$-wave solutions to be found in the three dimensional $(B,F,\E)$ solution space, which we consider to be prohibitively expensive computationally. 
By restricting only to profiles displaying a single trapped bubble, this solution space simplifies to a single branch throughout the $(B,F,\E)$ bifurcation diagram, which we projected to the $(B,F)$-plane for simplicity.

\subsection{Limiting pure-$1$ solution space}

\noindent  This branch of limiting solutions is displayed in figure~\ref{fig:NewSol2} $(a)$. 
These solutions were found from the same numerical procedure as in \S\ref{sec:num}.
An initial limiting solution, displaying one trapped bubble, is found by increasing $\E$.
The energy constraint is then replaced by a trapped bubble condition, which forces the second turning point of $X(\phi)$ to a value of $-0.5$.
We then explore the $(B,F)$-solution space by continuation.

We note that as $B \to \infty$ along this branch, the wave profile approaches the limiting pure capillary solution found by \cite{crapper1957exact} (see their Fig.~1) with an amplitude of $0.730$. Solution $(b)$, with $B=4.050$, is an example of this.
As $B \to 0$ along the same branch, the solution approaches a depressive solitary wave, demonstrated by solution $(d)$.
This is since the small $B$ limit is related to the solitary wave limit of $L_{\lambda} \to \infty$.
The solutions calculated by \cite{schwartz1979numerical} (see their Fig.~2) form the intermediate range between these two limits.

\bibliographystyle{jfm}

\end{document}